%% file: paper.tex
\documentclass[longauth]{aaEC}  

\makeatletter
\renewcommand*\aa@pageof{, page \thepage{} of \pageref*{LastPage}}
\makeatother

\usepackage{euclid}
\usepackage{graphicx}
\usepackage{natbib}
\usepackage{scalerel}
\usepackage{comment}
\usepackage{txfonts}
\usepackage{multirow}
\usepackage{multicol}
\usepackage{colortbl}
\usepackage{float}
\usepackage{array}
\usepackage[table]{xcolor}
\usepackage{soul}
\usepackage[pdfencoding=auto,psdextra]{hyperref}  
\usepackage{longtable}
\hypersetup{colorlinks=true,
    linkcolor=blue,
    filecolor=magenta,      
    urlcolor=blue,
    citecolor=blue}
\usepackage{color}
\usepackage[switch, modulo]{lineno}

\newcommand {\T}{Table\,}
\newcommand {\Sec}{Sect.\,}
\newcommand {\Fig}{Fig.\,}

\newcommand{\cc}[1]{{\color{blue}{#1}}}

\begin{document}


\title{Euclid Quick Data Release (Q1)}
\subtitle{The first catalogue of strong-lensing galaxy clusters}    

\include{authorlist}

\date{Received February 14, 2020; accepted February 14, 2020}


\abstract{
    We present the first catalogue of strong lensing galaxy clusters identified in the Euclid Quick Release 1 observations (covering $63.1\,\mathrm{deg^2}$). This catalogue is the result of the visual inspection of 1260 cluster fields. Each galaxy cluster was ranked with a probability, $\mathcal{P}_{\mathrm{lens}}$, based on the number and plausibility of the identified strong lensing features.
    Specifically, we identified 83 gravitational lenses with $\mathcal{P}_{\mathrm{lens}}>0.5$, of which 14 have $\mathcal{P}_{\mathrm{lens}}=1$, and clearly exhibiting secure strong lensing features, such as giant tangential and radial arcs, and multiple images. Considering the measured number density of lensing galaxy clusters, approximately $0.3\,\mathrm{deg}^{-2}$ for $\mathcal{P}_{\mathrm{lens}}>0.9$, we predict that \Euclid\ will likely see more than 4500 strong lensing clusters over the course of the mission. Notably, only three of the identified cluster-scale lenses had been previously observed from space. Thus, \Euclid has provided the first high-resolution imaging for the remaining $80$ galaxy cluster lenses, including those with the highest probability. The identified strong lensing features will be used for training deep-learning models for identifying gravitational arcs and multiple images automatically in \Euclid observations. This study confirms the huge potential of \Euclid for finding new strong lensing clusters, enabling exciting new discoveries on the nature of dark matter and dark energy and the study of the high-redshift Universe.
}


\titlerunning{The first catalogue of strong-lensing galaxy clusters}

\authorrunning{Euclid
Collaboration: P. Bergamini et al.}
   
\maketitle

\begin{figure}[]
\centering
\includegraphics[width=0.97\linewidth]{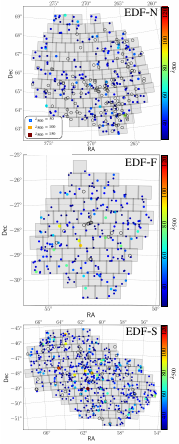}
\caption{Spatial distribution of galaxy clusters in the DLS and GCWG samples (see \Sec\ref{sec:GCcatalogue}). From top to bottom: three zoomed-in views of the Euclid Deep Field North (EDF-N), Euclid Deep Field Fornax (EDF-F), and Euclid Deep Field South (EDF-S) regions. Dot colours and sizes are scaled according to cluster richness, while empty circles mark clusters for which the richness value is not measured by \cite{Wen_2024}. A cut in richness at $\lambda_{500}>30$ is applied to the catalogue by \cite{Wen_2024} to select the galaxy clusters for visual inspection.}
\label{fig:sky}
\end{figure}

\section{Introduction}

Gravitational lensing in galaxy clusters is widely recognised as a powerful tool for determining their total mass distribution, including baryonic and dark components, across scales ranging from several kpc to Mpc. This can be achieved through strong lensing \citep[e.g.,][]{Grillo_2015, Mahler2018, Bergamini2023, Diego2023, Furtak2023, Acebron2024}, weak lensing \citep[e.g.,][]{Medezinski2016, Umetsu2020}, or combined analyses \citep[e.g.,][]{Jauzac2016, Liesenborgs2020, Niemiec2023}. These methods also allow for the measurement of key cosmological parameters that govern cosmic geometry and the expansion rate \citep[e.g.,][]{Jullo2010, Caminha2016, Caminha2022, Magana2018, Grillo2024}, as well as the detailed study of the intrinsic properties of background lensed galaxies \citep[e.g.,][]{Coe_2013, Bouwens_2014, Hashimoto_2018, Mestric2022, Vanzella2024}.

Unfortunately, strong-lensing galaxy clusters are relatively rare \citep[][]{Oguri_2010}, underscoring the critical role of wide-field surveys in building statistical samples. High-resolution optical and near-infrared imaging is particularly effective for identifying strong-lensing features with precision and reliability. However, such observations are typically constrained by small fields of view, often limited to a few arcminutes. This restriction has likely contributed to the relatively small number of studies leveraging high-resolution, space-based observations to significantly expand samples of strong-lensing galaxy clusters \citep[see, e.g., the MACS, beyond-MACS, or SGAS-HST surveys,][]{Ebeling_2001, Ebeling2024, Sharon2020}.

Enabled by its unique combination of high-resolution and wide-area survey capabilities, \Euclid\ is poised to revolutionise the field of strong lensing by galaxy clusters. Although it is scheduled to ultimately survey in the optical and near-infrared approximately $14\,000\,\mathrm{deg^2}$ of the sky \citep{EuclidSkyOverview}, the \cite{Q1cite} already holds immense potential to contribute significantly to the scientific goals outlined above. The Q1 dataset (\citealt{Q1-TP001, Q1-TP002, Q1-TP003, Q1-TP004, Q1-TP005, Q1-TP006, Q1-TP007}; \cc{Euclid Collaboration: Paterson et al. in prep.}) encompasses the three Euclid Deep Fields (EDFs), covering a total area of $63.1\,\mathrm{deg^2}$ to the depth of the Euclid Wide Survey \citep[EWS, ][]{Scaramella-EP1}.

This potential is exemplified by the first results from {\Euclid}'s Early Release Observations \citep[][]{Cuillandre2024} of two lensing galaxy clusters, Abell 2390 and Abell 2764 \citep[][\cc{Abriola et al. in prep.}; \cc{Diego et al. in prep.}]{Atek2024}. These studies highlight \Euclid ability to map the total mass distribution of such systems and to identify candidate high-redshift ($z > 6$) dropout sources \citep[see also][]{Weaver2024}. By opening a broader discovery window and providing a platform for validating detection and analysis techniques, the Q1 release marks an exciting milestone in the mission’s scientific journey.

This work aims to investigate {\Euclid}'s sheer potential to uncover lensing galaxy clusters and build the first catalogue of cluster-scale strong-lensing features and giant arcs in \Euclid. This effort is crucially needed to generate efficient training sets for sophisticated machine-learning algorithms to be applied to the \Euclid imaging data in the Data Release 1 (DR1) and beyond (e.g., \cc{Bazzanini et al. in prep.}), where analyses based on visual inspections alone will be extremely challenging. 
To achieve these goals, we perform the first systematic visual inspection in the Q1 data, over an effective area of $4.4\,\deg^2$, about 10\% of Q1, and study the prevalence of strong-lensing clusters in an optically-selected sample of galaxy clusters based on richness. Given the large areas to be examined, we exploit the tool \texttt{galaxyvote} developed by \cc{Meneghetti et al.} (\cc{in prep.}), which allows inspectors to efficiently scrutinise and go through large samples of images, with functionalities that are equally powerful to search for galaxy-, group-, and cluster-scale lenses.

This paper is organised as follows. In \Sec\ref{sec:Data}, we concisely describe the \Euclid imaging data and the creation of colour images of the candidate lensing galaxy clusters. Section \ref{sec:methodology} provides an overview of the considered galaxy cluster catalogue, the properties and capabilities of the newly developed tool, \texttt{galaxyvote}, and the adopted methodology for the visual inspection. Our results are presented and discussed in \Sec\ref{sec:results}. Finally, we draw our conclusions in \Sec\ref{sec:conclusions}. Throughout the paper, we assume a flat ${\Lambda\mathrm{CDM}}$ cosmology with $H_0 = 70$ $ \mathrm{km~s^{-1}~Mpc^{-1}}$, and matter density $\mathrm{\Omega_{m}}=0.3$. Magnitudes are given in the AB system \citep{Oke1974}. Statistical uncertainties are quoted as the 68\% confidence levels.

\section{Data and colour image generation}
\label{sec:Data}
All the results presented in this work are based on the \Euclid\ observations in the \IE, \YE, \JE, and \HE\ photometric filters \citep[][]{EuclidSkyVIS, EuclidSkyNISP} from the Q1 data release, which covers a total sky area of approximately $63.1\,\mathrm{deg^2}$, corresponding to the combined areas of the three EDFs (see \Fig\,\ref{fig:sky}, EDF-N $22.9\,\mathrm{deg^2}$, EDF-F $12.1\,\mathrm{deg^2}$, and EDF-S $28.1\,\mathrm{deg^2}$) but observed at the depth of the Euclid Wide Survey (EWS). For the \IE\ filter, this corresponds to a signal-to-noise ratio $\mathrm{S/N} \geq 10$ for extended sources with a full width at half maximum (FWHM) of $\ang{;;0.3}$ and AB magnitude of 24.5 within an aperture with a diameter of $\ang{;;1.3}$. Instead, the observations in the near-IR bands reach a $\mathrm{S/N} \geq 5$ for point sources with AB magnitude of 24.0 \citep{EuclidSkyOverview}. The \IE, \YE, \JE, and \HE\ images have a spatial sampling of \ang{;;0.1} per pixel, equivalent to the native pixel scale of the \IE\ images, and a point-spread function of \ang{;;0.13}, \ang{;;0.33},
\ang{;;0.35}, and \ang{;;0.36} FWHM, respectively \citep{EuclidSkyOverview, Laureijs_2011}. We note that although the original spatial sampling of the \YE, \JE, and \HE\ images is \ang{;;0.3} per pixel, it is oversampled to match the \IE\ pixel scale by the \Euclid\ data-reduction pipeline.

We create colour images by combining the \Euclid\ observations in different bands using the following method. 
To begin, we use the {\tt STIFF} software \citep{2012ASPC..461..263B} to create colour images in the Red-Green-Blue (RGB) colour space. Specifically, we use the \HE, \YE, and \IE\ bands for the red, green, and blue channels, respectively. We use the automatic sky background intensity and colour balance implemented in {\tt STIFF}. At the same time, we manually adjust the contrast and brightness to provide satisfactory visibility for low-surface-brightness sources.

The spatial resolution of these images is penalised by the equal weight given to the \IE\ and near-IR bands. To improve the perceived spatial resolution, we perform a two-step image manipulation. First, we map the images from the RGB to the CIELAB colour space. The latter expresses colour as three values: {\em L} for perceptual lightness, and {\em a} and {\em b} for the four unique colours of human vision, red, green, blue, and yellow. Then, we substitute the {\em L} channel with the image in the \IE\ band. Finally, we map the image back to RGB colour space. The images are saved in the well-known Tag Image File Format (TIF).

This process results in images whose spatial resolution is driven by the \IE\ band, while the colour information is retained from the combination of \HE, \YE, and \IE\ observations. Thus, the visibility of small-scale morphological features in the sources is greatly enhanced. For example, compact star-forming regions in distant galaxies are much better resolved in the \IE\ than in the near-IR bands and, therefore, they are more clearly visible in the colour images created with the procedure outlined above than in the RGB image initially created with {\tt STIFF}.

It is important to note that the methodology employed to prepare the colour images is often used in astrophotography for visualisation purposes only. These images are unsuitable for scientific analysis (e.g., to perform multi-band photometry or measure photometric redshifts). However, the purpose of our experiment is to identify strong lensing features visually. This task is facilitated by the higher resolution and colour contrast achieved with the image processing that we have outlined. An additional caveat is that using the \IE\ band for the lightness channel dims the features only visible in the near-IR bands. For example, \IE-dropout sources appear strongly attenuated in our colour images.     

\begin{figure}[]
\centering
\includegraphics[width=\linewidth]{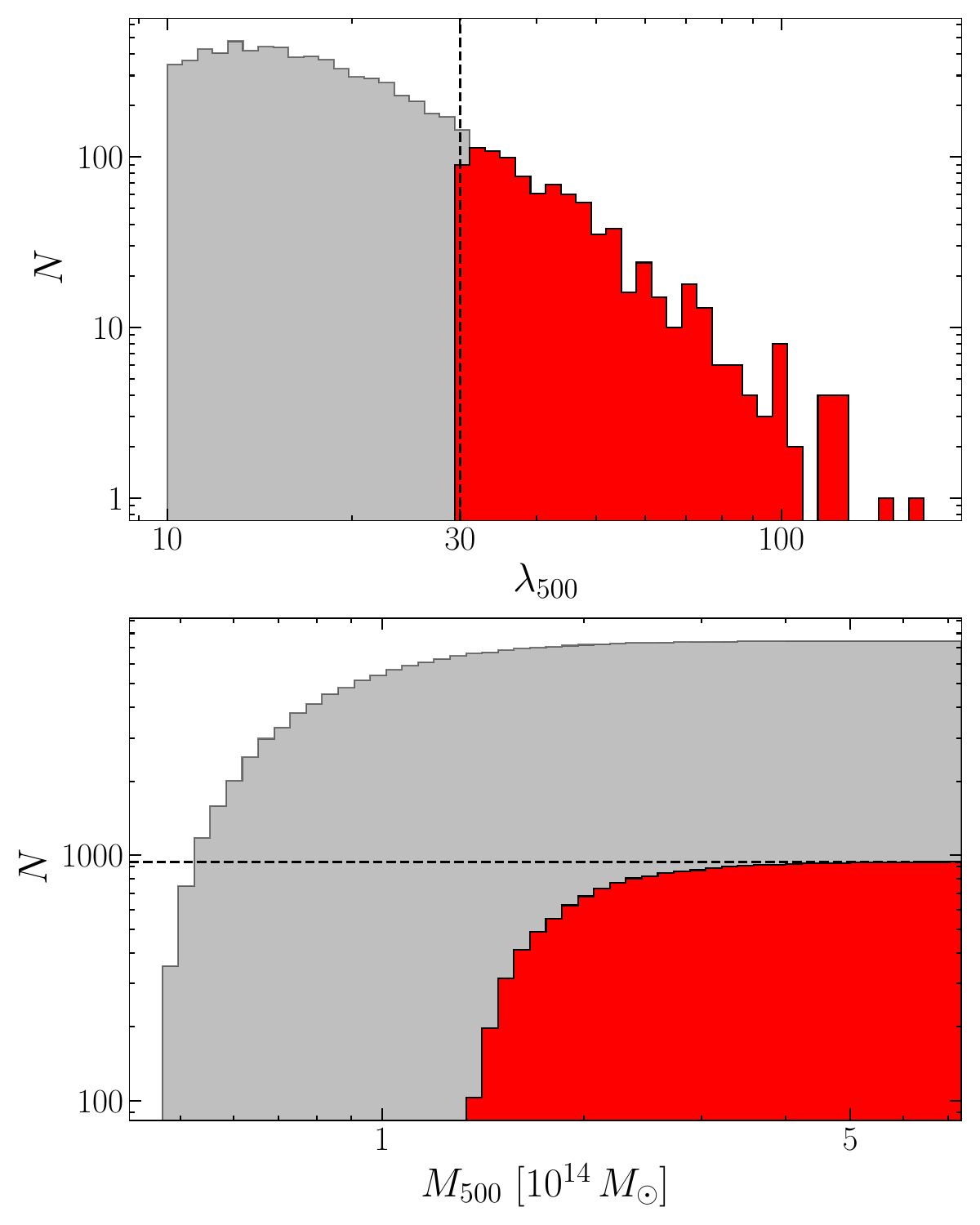}
\caption{Richness and mass distributions of  galaxy clusters in Q1 from the catalogue by \cite{Wen_2024}. 
The top panel shows the distribution of cluster richness values ($\lambda_{500}$), while the bottom panel presents the cumulative distribution of cluster total masses ($M_{500}$). Grey histograms represent the galaxy clusters in Q1 detected in the DESI Legacy Imaging Surveys by \cite{Wen_2024}, whereas the red distributions correspond to the 939 clusters with $\lambda_{500}>30$. 
}
\label{fig:histoM500}
\end{figure}

\section{Methodology}
\label{sec:methodology}

In this section, we describe the sample of visually-inspected candidate lens galaxy clusters (hereafter, the VI catalogue), which is created by combining two main galaxy cluster catalogues (see \Sec \ref{sec:GCcatalogue}). Section \ref{sec:VI_tool} presents the new visual inspection tool, purposely built for this project. The visual inspection methodology is detailed in \Sec \ref{sec:VI}.

\begin{figure}[h!]
\centering
\includegraphics[width=0.9\linewidth]{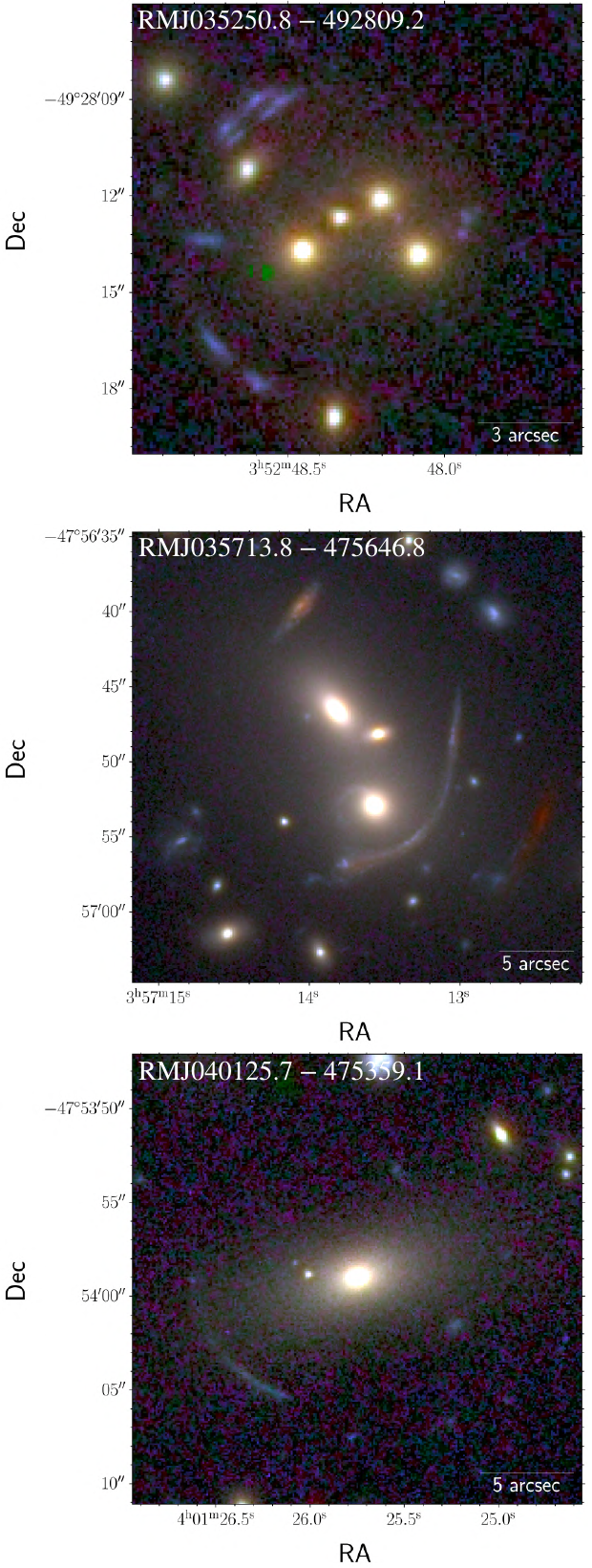}
\caption{The three secure galaxy cluster strong gravitational lenses ($\mathcal{P}_{\mathrm{lens}}=1$) identified in the calibration phase (GCWG sample).}
\label{fig:clusters_phase1}
\end{figure}

\subsection{Galaxy cluster catalogue}
\label{sec:GCcatalogue}
To carry out the proposed experiment, we first considered the galaxy cluster catalogue presented by \citet{Wen_2024}, based on multi-band photometry from the DESI Legacy Imaging Surveys Data Release (DR) 9 and 10 \citep{Dey_2019}, and the Wide-field Infrared Survey Explorer \citep[WISE, ][]{Wright_2010}. In summary, four optical bands ($g$, $r$, $i$, and $z$) and two mid-infrared bands (W1 and W2) were used. The photometric information was complemented with available spectroscopy from the Two Micron All-Sky Survey \citep[2MASS,][]{Huchra_2012},
 the Sloan Digital Sky Survey DR17 \citep[SDSS DR17,][]{SDSS_2022},
 and the Dark Energy Spectroscopic Instrument Early Data Release \citep[DESI EDR,][]{DESI_2024}. 
Galaxy clusters were identified as overdensities of stellar mass in  cluster galaxies centred on the brightest cluster galaxy (BCG) candidates within a redshift bin of $\Delta z= 0.04\,(1+z)$ for $z \leq 0.7$ or $\Delta z= 0.15\, z - 0.037$ for $z > 0.7$. BCGs are identified based on their $rz$W1 magnitudes, estimated redshift values, and measured stellar masses $M_{\ast} \geq 10^{11} M_{\odot}$.
The publicly available catalogue lists about 1.58 million candidate galaxy clusters, of which 7446 are located within the Q1 footprint. We then applied a cut in richness $\mathrm{\lambda}_{500}>30$, resulting in 939 galaxy clusters (hereafter, DLS galaxy clusters). As shown in \Fig\ref{fig:histoM500}, this is equivalent to applying a cut of $M_{500} \gtrsim 1.36 \times 10^{14} M_{\odot}$.
The spatial distribution of the DLS galaxy clusters in the three Q1 fields is shown in \Fig \ref{fig:sky}, colour-coded according to the estimated value of $\lambda_{500}$ by \citet{Wen_2024}.

As discussed in \citet{Wen_2024}, while their photometric selection has good completeness ($80\%\! -\! 94\%$, depending on the survey considered for the comparison), some galaxy clusters are not identified.
Thus, we complement the DLS galaxy cluster sample with a compilation created by merging 15 catalogues of known galaxy clusters (courtesy of Jean-Baptiste Melin, see also \cc{Euclid Collaboration: Bhargava et al. in prep.}). In detail, it includes galaxy clusters visible in the Q1 and DR1 footprints from the following catalogues (the number of Q1 systems is given in parentheses): the Meta-Catalogue of the compiled properties of X-ray detected Clusters of galaxies \citep[MCXC, ][35 galaxy clusters]{Piffaretti_2011, Sadibekova_2024}; the extended ROentgen Survey with an Imaging Telescope Array (eROSITA) cluster catalogue \citep[][125 galaxy clusters]{Bulbul_2024, Kluge_2024}; the Meta Catalogue of SZ clusters (MCSZ, see \cc{Euclid Collaboration: Bhargava et al. in prep.}, 28 galaxy clusters); the Combined {\it Planck}-RASS catalogue of X-ray–SZ sources \citep[ComPRASS, ][six galaxy clusters]{Tarrio_2019}; the Meta-Catalogue of Cluster Dispersions (MCCD, see \cc{Euclid Collaboration: Bhargava et al. in prep.}, two galaxy clusters); the RedMaPPer galaxy cluster catalogue from the Dark Energy Survey science verification data \citep[RM DES, ][177 galaxy clusters]{Rykoff_2016, Abbott_2020}; the catalogue of clusters of galaxies identified from the Sloan Digital Sky Survey by \citet[][WHL SDSS, 142 galaxy clusters]{Wen_2012}; the Abell catalogue of rich clusters \citep[][23 galaxy clusters]{Abell_1958, Abell_1989}; the MARD-Y3 catalogue \citep[][19 galaxy clusters]{Klein_2019}; the XMM CLuster Archive Super Survey catalogue \citep[X-CLASS, ][seven galaxy clusters]{Koulouridis_2021}; the PSZSPT catalogue \citep[][10 galaxy clusters]{Melin_2021}; the PSZ-MCMF catalogue of {\it Planck} clusters over the DES region \citep[][seven galaxy clusters]{Hernandez-Lang_2023}; the SPT-SZ MCMF cluster catalogue \citep[][13 galaxy clusters]{Klein_2024a}; the RASS-MCMF cluster catalogue \citep[][44 galaxy clusters]{Klein_2023}; and the ACT-DR5-MCMF galaxy cluster catalogue \citep[][21 clusters]{Klein_2024b}. From this sample, 484 unique galaxy clusters (i.e., those at least \ang{;1.5;} apart) fall within the Q1 footprint, of which 117 are not included in \citet{Wen_2024}. We label this sample as GCWG (galaxy cluster working group) sample. The spatial distribution of these galaxy clusters in the three Q1 fields, marked as empty black circles, is shown in \Fig \ref{fig:sky}.

In summary, to build the sample of galaxy clusters to be considered for visual inspection (the VI catalogue), we integrated the 939 DLS systems with the GCWG sample, resulting in a final catalogue of 1056 candidate lens clusters. Additionally, we removed 10 galaxy clusters from the DLS catalogue with projected separations smaller than \ang{;1.5;}. These occurrences are, in fact, not duplicate candidate galaxy clusters but systems lying angularly close but at different redshifts. Since the \Euclid image cutouts adopted for the visual inspection have a size of $\ang{;4;} \times \ang{;4;}$ (see \Sec \ref{sec:Data}), this entails that both systems are visually inspected within a single cutout. Finally, we excluded 220 galaxy clusters located close to the boundaries of the Q1 tiles, whose cutouts contain more than $50\%$ of null values. 
The final number of cutouts to visually inspect is then equal to 826. 

\begin{figure}[t!]
\centering
\includegraphics[width=\linewidth]{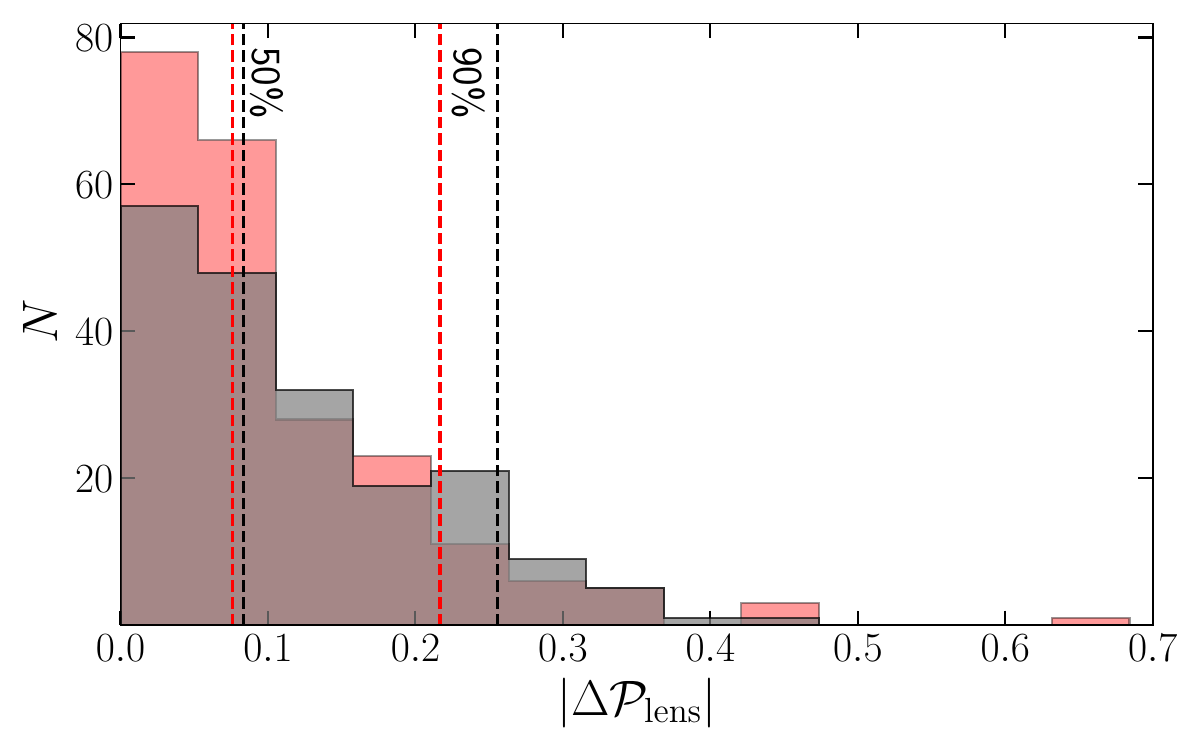}
\caption{Difference between the $\mathcal{P}_{\mathrm{lens}}$ values, $\Delta\mathcal{P}_{\mathrm{lens}}$, attributed to the same clusters by different groups of inspectors and/or at different times. The black distribution shows the $\Delta\mathcal{P}_{\mathrm{lens}}$ obtained for different occurrences of the same cluster in the GCWG sample during the calibration phase of the visual inspection. The red distribution refers to the $\Delta\mathcal{P}_{\mathrm{lens}}$ attributed to the same cluster during the two phases of the visual inspection. The black and red dashed vertical lines mark the 50th and 90th percentiles of the $\mathcal{P}_{\mathrm{lens}}$ black and red distributions, respectively.}
\label{fig:duplicate_clusters}
\end{figure}

\subsection{The \texttt{galaxyvote} platform}
\label{sec:VI_tool}
For the visual inspection of the \Euclid images, we use the \texttt{galaxyvote} web application (\cc{Meneghetti et al. in prep.}). In this section, we provide a short description of its functionalities. 

The application back-end uses the {\tt Flask} framework\footnote{\url{https://flask.palletsprojects.com/en/stable/}} to define routes (URLs) for the web application and associate them with \texttt{Python} functions. These functions are designed to interact with an application database. This is a relational database containing several tables. The software can handle multiple experiments using different image collections, providing access to many users, and managing assignments. Additional tables enable the storing of grades, comments, and even geometrical figures drawn by the users to mark specific regions of the images.
The application front-end is based on HTML and \texttt{JavaScript}. It employs the \texttt{Bootstrap} \texttt{CSS} framework\footnote{\url{https://getbootstrap.com/}} to quickly and efficiently design and style responsive and modern web pages. 

For a given visual inspection experiment, a configuration script allows us to initialise the application database, populating it with the list of images and authorised users. These users can log in using pre-assigned credentials, accessing their dedicated workspaces. 
Administrators set the number of users to evaluate each image during the configuration. The system optimally creates the assignments based on the number of users and images. Each user's workspace contains a gallery of assigned images.

The users inspect the \Euclid images through a web page containing the following features.
\begin{itemize}
\item An $800\times800$ pixel viewer allows the visualisation of the images. We use the javascript library {\tt OpenSeaDragon} for smooth image navigation, including mouse-enabled zoom in and out functions. Users can opt to enter full-screen mode to improve their visualisation experience. To speed up the image panning and scrolling, we tile the images, converting them in the pyramidal Deep-Zoom-Image (DZI) format.\footnote{\url{https://shorturl.at/oZe52}} {\tt OpenSeaDragon} supports several image-serving protocols for tiled sources, including DZI.
\item Basic filters, such as increasing and reducing contrast and brightness to enable users to improve the visibility of specific sources in the images. These filters are implemented using the {\tt OpenSeaDragonFiltering} add-on.\footnote{\url{https://github.com/usnistgov/OpenSeadragonFiltering}}
\item A rectangle drawing tool. Users can enable drawing rectangles in the image viewer. They can draw multiple rectangles to mark the positions of interesting features in the image and save them into the database. For our experiment, we ask users to draw rectangles on gravitational arcs and arclets, and sets of multiple images of background sources.
\item A voting panel. Users are asked to classify the inspected image. For our experiment, we set up three classes, namely: `Certain Lens' (A), if the image contains any features that the user associates with a strong lensing effect with extremely high confidence; `Probable Lens' (B) if the detected features lead the user to believe that there is a strong lensing effect, but with lower confidence; and `No Lens' (C) if the user does not detect any strong lensing feature.
\item A comment box. Users can leave feedback on their feature identifications. In particular, for our experiment, they were instructed to explain why they believed that strong lensing effects could explain the detected features. In addition, they could describe their identifications in more detail if they wanted.   
\end{itemize} 

\begin{figure}[t]
\centering
\includegraphics[width=\linewidth]{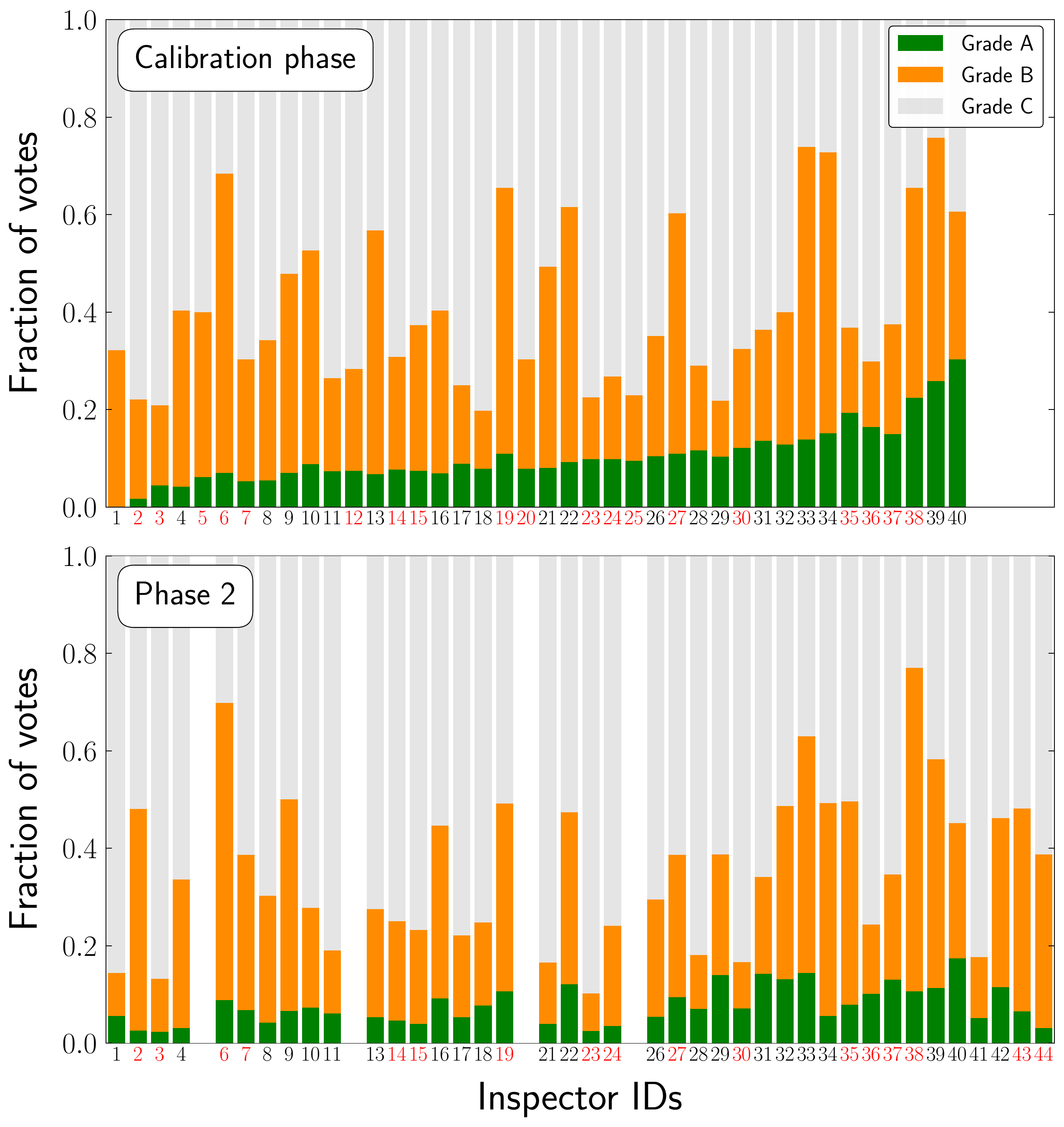}
\caption{Fraction of A, B, and C grades assigned by each inspector, identified by unique numerical IDs, to the candidate cluster lenses. The results from the calibration phase and from Phase 2 are shown in the upper and lower panels, respectively. Inspectors with little or no experience in strong gravitational lensing by galaxy clusters are marked with red IDs. Inspectors who graded less than two clusters are not included.}
\label{fig:users}
\end{figure}

\begin{figure*}[]
\centering
\includegraphics[width=0.95\linewidth]{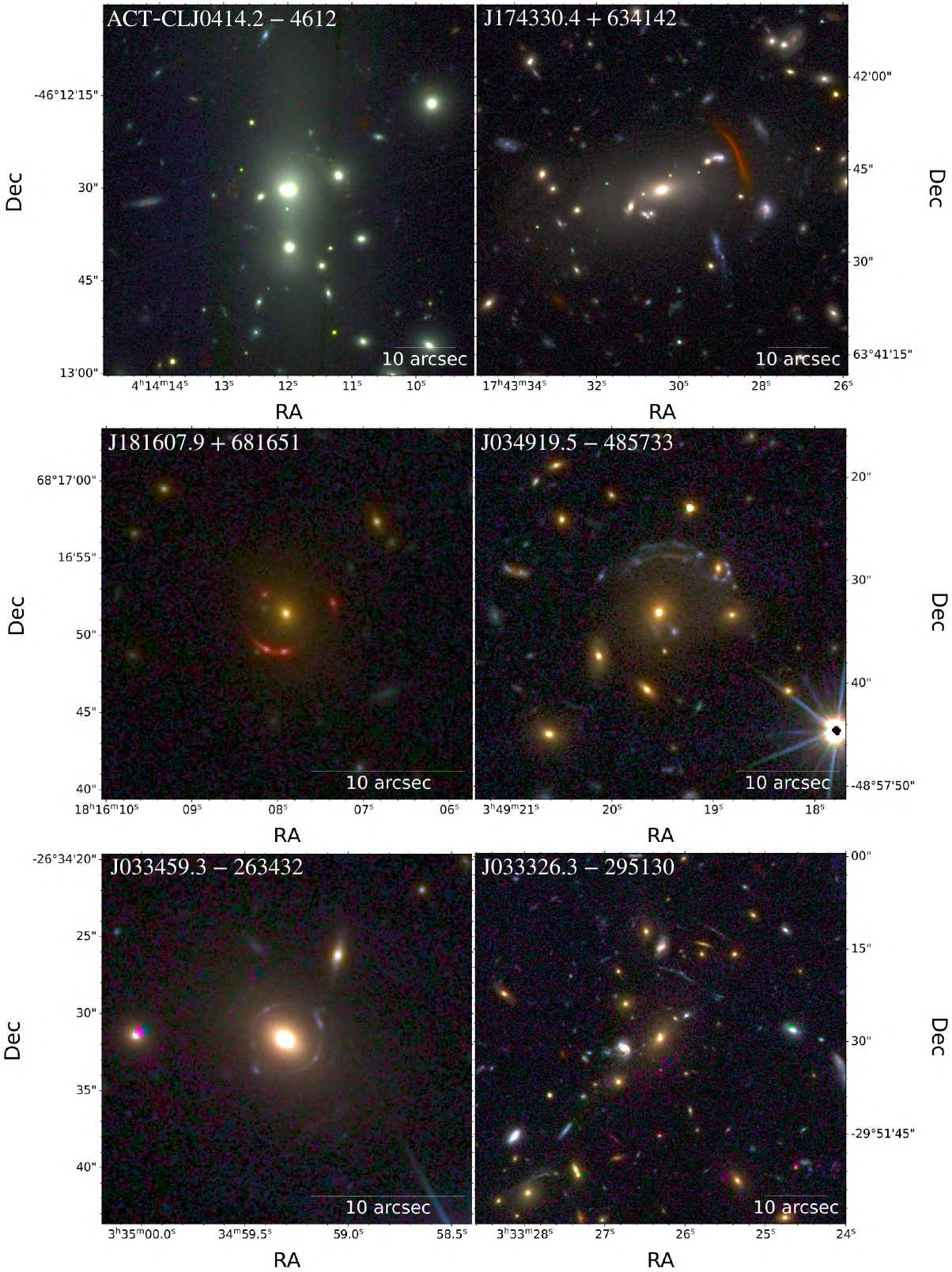}
\caption{Six out of the 13 secure galaxy cluster strong gravitational lenses ($\mathcal{P}_{\mathrm{lens}}=1$) identified in the Phase 2 run (VI catalogue). The remaining seven secure Phase 2 lenses are shown in Figs.\,\ref{fig:VIgradeA2}, \ref{fig:bullet}, and \ref{fig:clusters_phase1} (RMJ035713.8$-$475646.8 and RMJ040125.7$-$475359.1).} 
\label{fig:VIgradeA1}
\end{figure*}

\begin{figure*}[]
\centering
\includegraphics[width=0.95\linewidth]{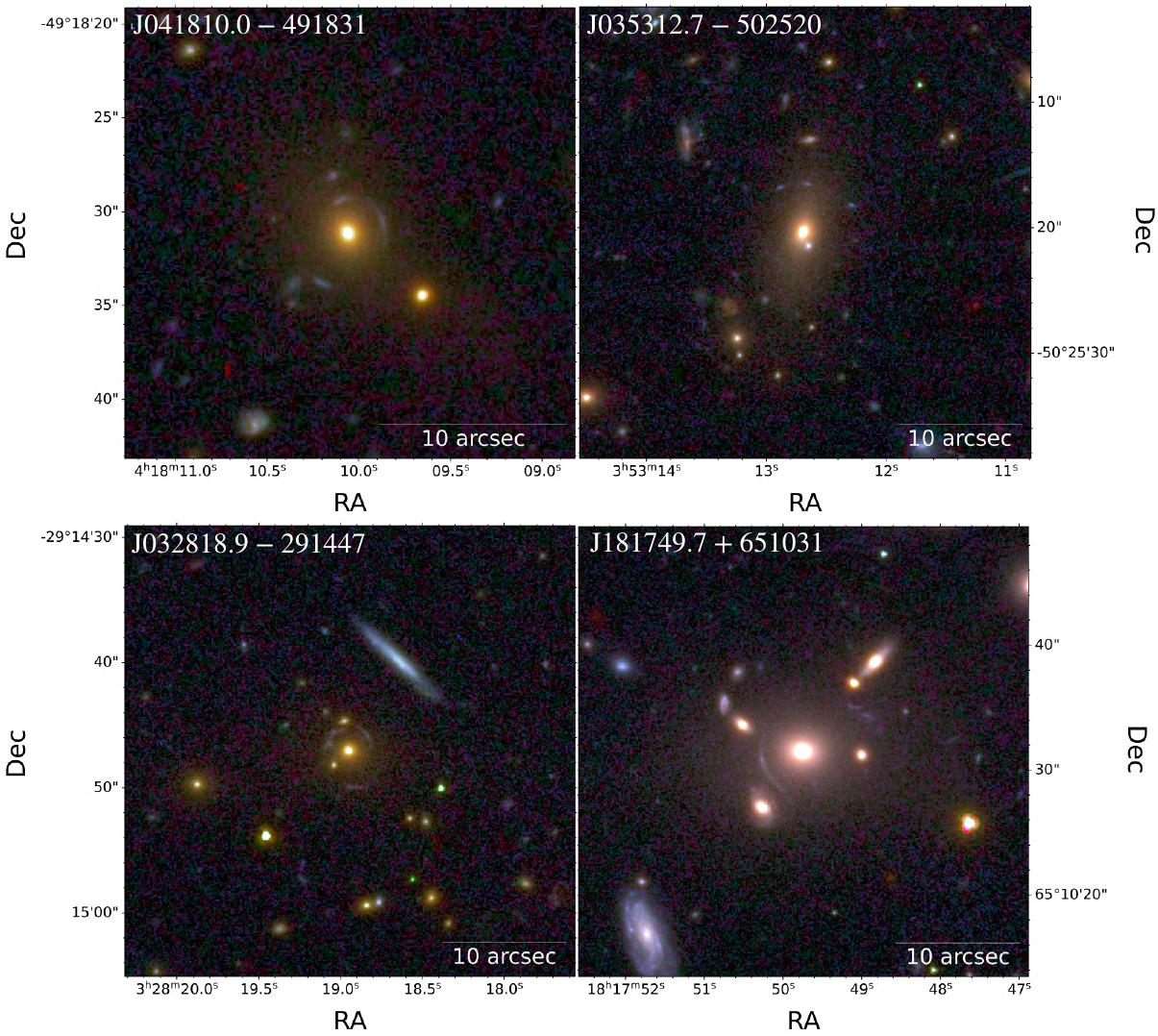}
\caption{Four out of the 13 secure galaxy cluster strong gravitational lenses ($\mathcal{P}_{\mathrm{lens}}=1$) identified in the Phase 2 run (VI catalogue). The remaining nine secure Phase 2 lenses are shown in Figs.\,\ref{fig:VIgradeA1}, \ref{fig:bullet}, and \ref{fig:clusters_phase1} (RMJ035713.8$-$475646.8 and RMJ040125.7$-$475359.1).}
\label{fig:VIgradeA2}
\end{figure*}

\begin{table*}[h!]
\caption{List of galaxy clusters in the calibration phase (GCWG sample) with a probability $\mathcal{P}_{\mathrm{lens}}>0.5$ of being a strong gravitational lens. For each galaxy cluster, we report the total number of A, B, and C grades ($N_\mathrm{A}$, $N_\mathrm{B}$, and $N_\mathrm{C}$) assigned by the inspectors. The star ($\star$) and dagger ($\dagger$) symbols mark the gravitational lenses previously observed by the {\it Hubble} Space Telescope and {\it James Webb} Space Telescope, respectively. The
catalogue is published in its entirety in machine-readable format. A portion listing the secure gravitational lenses ($\mathcal{P}_{\mathrm{lens}}=1$) is shown here for guidance regarding its form and content.}              
\label{table:phase1_lenses}      
\centering  
\begin{tabular}{lcccccc}
\hline
\hline
Name & RA & Dec & $N_\mathrm{A}$ & $N_\mathrm{B}$ & $N_\mathrm{C}$ & $\mathcal{P}_{\mathrm{lens}}$ \\
\hline
\\
RMJ040125.7$-$475359.1 & 60.35721 & $-$47.89974 & 4 & 0 & 0 & 1.00 \\
RMJ035713.8$-$475646.8 & 59.30752 & $-$47.94633 & 14 & 0 & 0 & 1.00 \\
RMJ035250.8$-$492809.2 & 58.21182 & $-$49.46921 & 6 & 0 & 0 & 1.00 \\
... & ... & ... & ... & ... & ... & ... \\
\hline
\end{tabular}
\end{table*}

\begin{table*}[h!]
\caption{List of galaxy clusters in the Phase 2 run (VI catalogue) ranked with a probability, $\mathcal{P}_{\mathrm{lens}}$, greater than 0.5 to be strong gravitational lenses. For each galaxy cluster, we report the number of A, B, and C votes ($N_\mathrm{A}$, $N_\mathrm{B}$, and $N_\mathrm{C}$) assigned by the inspectors. The star ($\star$) and dagger ($\dagger$) symbols mark the gravitational lenses previously observed by the {\it Hubble} Space Telescope and {\it James Webb} Space Telescope, respectively. The
catalogue is published in its entirety in machine-readable format. A portion listing the secure gravitational lenses ($\mathcal{P}_{\mathrm{lens}}=1$) is shown here for guidance regarding its form and content.}           
\label{table:phase2_lenses} 
\centering  
\begin{tabular}{lcccccc}
\hline
\hline
Name & RA & Dec & $N_\mathrm{A}$ & $N_\mathrm{B}$ & $N_\mathrm{C}$ & $\mathcal{P}_{\mathrm{lens}}$ \\
\hline
\\
ACT-CLJ0411.2$-$4819 & 62.80204 & $-$48.31183 & 7 & 0 & 0 & 1.00 \\
J174330.4+634142 & 265.87665 & 63.69491 & 7 & 0 & 0 & 1.00 \\
ACT-CLJ0414.2$-$4612 & 63.55490 & $-$46.20033 & 6 & 0 & 0 & 1.00 \\
RMJ035713.8$-$475646.8 & 59.30752 & $-$47.94633 & 6 & 0 & 0 & 1.00 \\
J034919.5$-$485733 & 57.33133 & $-$48.95924 & 6 & 0 & 0 & 1.00 \\
J181749.7+651031 & 274.45724 & 65.17541 & 6 & 0 & 0 & 1.00 \\
J181607.9+681651 & 274.03287 & 68.28094 & 6 & 0 & 0 & 1.00 \\
J041810.0$-$491831 & 64.54186 & $-$49.30865 & 6 & 0 & 0 & 1.00 \\
J035312.7$-$502520 & 58.30281 & $-$50.42233 & 7 & 0 & 0 & 1.00 \\
J033326.3$-$295130 & 53.35960 & $-$29.85820 & 7 & 0 & 0 & 1.00 \\
RMJ040125.7$-$475359.1 & 60.35721 & $-$47.89974 & 7 & 0 & 0 & 1.00 \\
J033459.3$-$263432 & 53.74707 & $-$26.57546 & 7 & 0 & 0 & 1.00 \\
J032818.9$-$291447 & 52.07892 & $-$29.24641 & 7 & 0 & 0 & 1.00 \\
... & ... & ... & ... & ... & ... & ... \\
\hline
\end{tabular}
\end{table*}

\subsection{The visual inspection}
\label{sec:VI}
While the primary aim of this experiment is to uncover new lens galaxy clusters, all inspectors were instructed to vote for any strong-lensing feature, regardless of whether it occurs in galaxy-, group-, or cluster-scale systems. The team of inspectors comprised 44 members of the Euclid Consortium (hereafter referred to as `experts' or `inspectors'), approximately half of whom are  specialists in strong lensing by galaxy clusters, and thus very familiar with cluster-scale strong lensing features. 

The visual inspection experiment is divided in two stages, as described below. We note that the experts did not have access to information about the galaxy cluster's properties, such as coordinates, redshift, richness, or total mass, during the entire visual inspection process, thus ensuring a blind search for strong-lensing features.

The first stage consists of a `calibration phase' in which the 44 inspectors were tasked to vote exclusively on a subsample of galaxy clusters contained in the GCWG sample (a total of 434 cutouts, see \Sec\,\ref{sec:GCcatalogue}). This collection of candidates is large enough to train the inspectors for the second phase. In fact, the number of assignments for each inspector varied between 55 and 87 so that each cutout could receive seven votes. This number of votes was determined to be the optimal compromise between the total number of required votes and the need for a robust final grade \citep[see e.g.,][]{Rojas_2023, Schuldt_2025}. 
The primary aim of this phase is to train the inspectors in identifying cluster-scale lenses and to familiarise them with the visual inspection tool. Additionally, this phase allows for identifying possible software bugs while implementing feedback on potential improvements for the second, larger stage. We note that duplicate entries in the GCWG sample were not removed during this phase, allowing us to evaluate how the same object is ranked by different groups of inspectors or by the same inspector at different times.
The calibration phase was completed by 40 out of the 44 experts involved, gathering 88\% of the expected number of votes in about a week. 

In contrast, the second stage examined the 826 \Euclid\ images in the complete VI catalogue presented in \Sec\,\ref{sec:GCcatalogue}. This phase involved 44 experts, of which 38 completed the inspection in about three weeks. Thus, we collected 90\% of the expected grades. As in the calibration phase we aimed at collecting seven grades per image. The number of assignments per inspector varied between 113 and 163.

\section{Results}
\label{sec:results}
In each phase, a probability value of being a strong gravitational lens, $\mathcal{P}_{\mathrm{lens}}$,  is assigned to each cluster. This probability is computed as
\begin{equation}
\label{Eq,:lensing_probability}
    \mathcal{P}_{\mathrm{lens}} = \frac{N_{\rm A}+0.5\, N_{\rm B}}{N_{\rm A}+N_{\rm B}+N_{\rm C}}\,,
\end{equation}
where $N_{\rm A}$, $N_{\rm B}$, and $N_{\rm C}$ are the numbers of A, B, and C grades given to the image, respectively.

A value of $\mathcal{P}_{\mathrm{lens}}=1$ is assigned only if all inspectors evaluate the cluster as a secure gravitational lens (all A grades). On the contrary, a value of $\mathcal{P}_{\mathrm{lens}}=0.5$ indicates that all inspectors are uncertain about the presence of strong lensing features (all B grades) and/or that the number of A grades (secure presence of strong lensing features) is equal to the number of C grades (secure absence of strong lensing features). In the following two sections, which describe the main results obtained from the two phases of the visual inspection, we consider clusters with $\mathcal{P}_{\mathrm{lens}}>0.5$ as potential strong gravitational lenses. 

\subsection{Calibration phase}
\label{sec:Phase1}
The calibration phase returned 25 strong lensing clusters. They are listed in \T\,\ref{table:phase1_lenses}. In \Fig\ref{fig:clusters_phase1}, we show the three cluster lenses with $\mathcal{P}_{\mathrm{lens}}=1$. As evident from the table, there are significant discrepancies in the total number of votes, $N=N_\mathrm{A}+N_\mathrm{B}+N_\mathrm{C}$, assigned to some of the clusters. For instance, the two clusters ACT-CLJ0411.2$-$4819 and RMJ040125.7$-$475359.1 have $N$ equal to 49 and 4, respectively. As previously mentioned, these differences arise from the presence of 60 duplicate clusters that appear at least twice in the GCWG sample. Thus, their images received a larger number of grades than in the case of clusters without duplicates.  
To test the robustness and self-consistency of the visual inspection methodology, we plot the differences between the $\mathcal{P}_{\mathrm{lens}}$ values assigned to different occurrences of the same clusters as a grey histogram in \Fig\ref{fig:duplicate_clusters}. For more than 50\% (90\%) of the duplicates, the difference in $\mathcal{P}_{\mathrm{lens}}$ is less than $0.09$ ($0.26$). This analysis demonstrates that the strong lensing probability is only marginally affected by variations in the group of inspectors voting for a cluster or by an inspector's vote assigned to a cluster at different times. As an additional test, we present in \Fig\ref{fig:users}, the percentage of A, B, and C votes assigned by each inspector to the visually inspected clusters. Inspectors with no prior experience in strong gravitational lensing by galaxy clusters are marked with red IDs on the horizontal axis. This analysis reveals the absence of any significant correlation between the number of votes and the users' level of expertise. The observed fluctuations are, in fact, consistent with the variance in number of cluster lenses within the cluster samples inspected by the users. 
The calibration phase did not reveal any major bugs in the visual inspection tool described in \Sec\,\ref{sec:VI_tool}, which is therefore also adopted for the second phase.

\begin{figure*}[h!]
\centering
\includegraphics[width=0.95\linewidth]{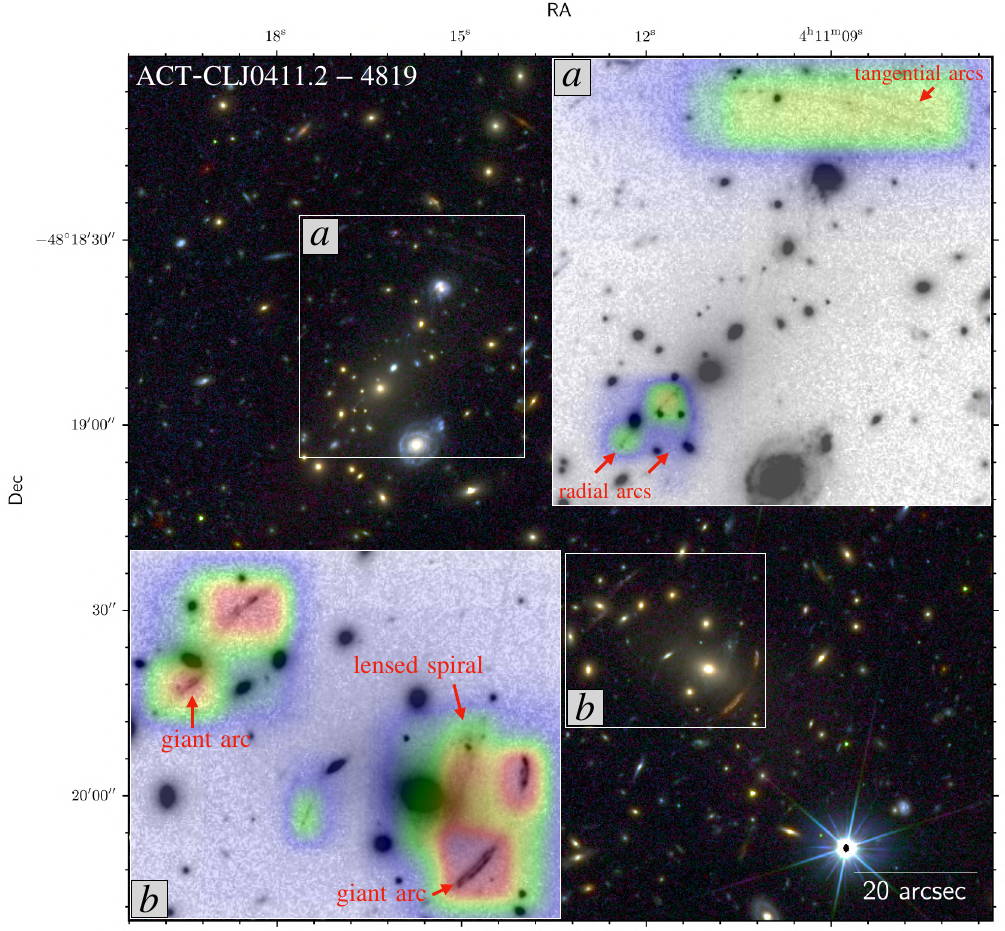}
\caption{Merging galaxy cluster identified as a secure gravitational lens in the second phase of the visual inspection. The zoom-ins centred on the two merging massive structures are superimposed, showing heat maps derived from the rectangular regions drawn by the inspectors during the visual inspection process. These maps highlight various strong lensing features, including tangential and radial arcs, as well as multiple images, with redder colours indicating more evident features.}
\label{fig:bullet}
\end{figure*}

\subsection{Phase 2}
\label{sec:Phase2}
In \T\,\ref{table:phase2_lenses}, we present the 76 strong gravitational lensing clusters identified during the second phase of visual inspection from the VI catalogue. In this phase, the inspectors identified 13 secure cluster lenses ($\mathcal{P}_{\mathrm{lens}}$=1), four of which were already found in the calibration phase (two were classified as secure lenses in both phases, while the other two were ranked with $\mathcal{P}_{\mathrm{lens}}$=0.99 and $\mathcal{P}_{\mathrm{lens}}$=0.83 in the calibration phase). However, the galaxy cluster RMJ035250.8$-$492809.2, identified as a secure lens in the first phase, is absent from the VI catalogue due to the richness cut of $\lambda_{500}>30$ applied during its creation. Specifically, \cite{Wen_2024} estimated a richness value of $\lambda_{500}=28.61$ for this cluster. In Figs.\,\ref{fig:VIgradeA1}, \ref{fig:VIgradeA2}, and \ref{fig:bullet}, we show the additional 11 secure cluster lenses identified in the second phase. 

Since 211 of the galaxy clusters visually inspected in the calibration phase were also re-inspected in the Phase 2, we show the absolute differences between the $\mathcal{P}_{\mathrm{lens}}$ values assigned to the same clusters in the two phases as a red histogram in \Fig\ref{fig:duplicate_clusters}. 
This additional test demonstrates that most clusters were consistently ranked between the two visual inspection phases. Specifically, 90\% of the analysed objects exhibit $\mathcal{P}_{\mathrm{lens}}$ differences of less than 0.22. Similarly to the calibration phase, in \Fig\ref{fig:users} we present the percentage of A, B, and C votes assigned to the clusters visually inspected by each user. As in the calibration phase, Phase 2 also does not reveal any significant bias in the distribution of votes across the sample of inspectors or as a function of user expertise.

\subsection{Characterisation of the physical properties of the lens clusters}
\label{sec:Clusterproperties}

\begin{figure}[t]
\centering
\includegraphics[width=\linewidth]{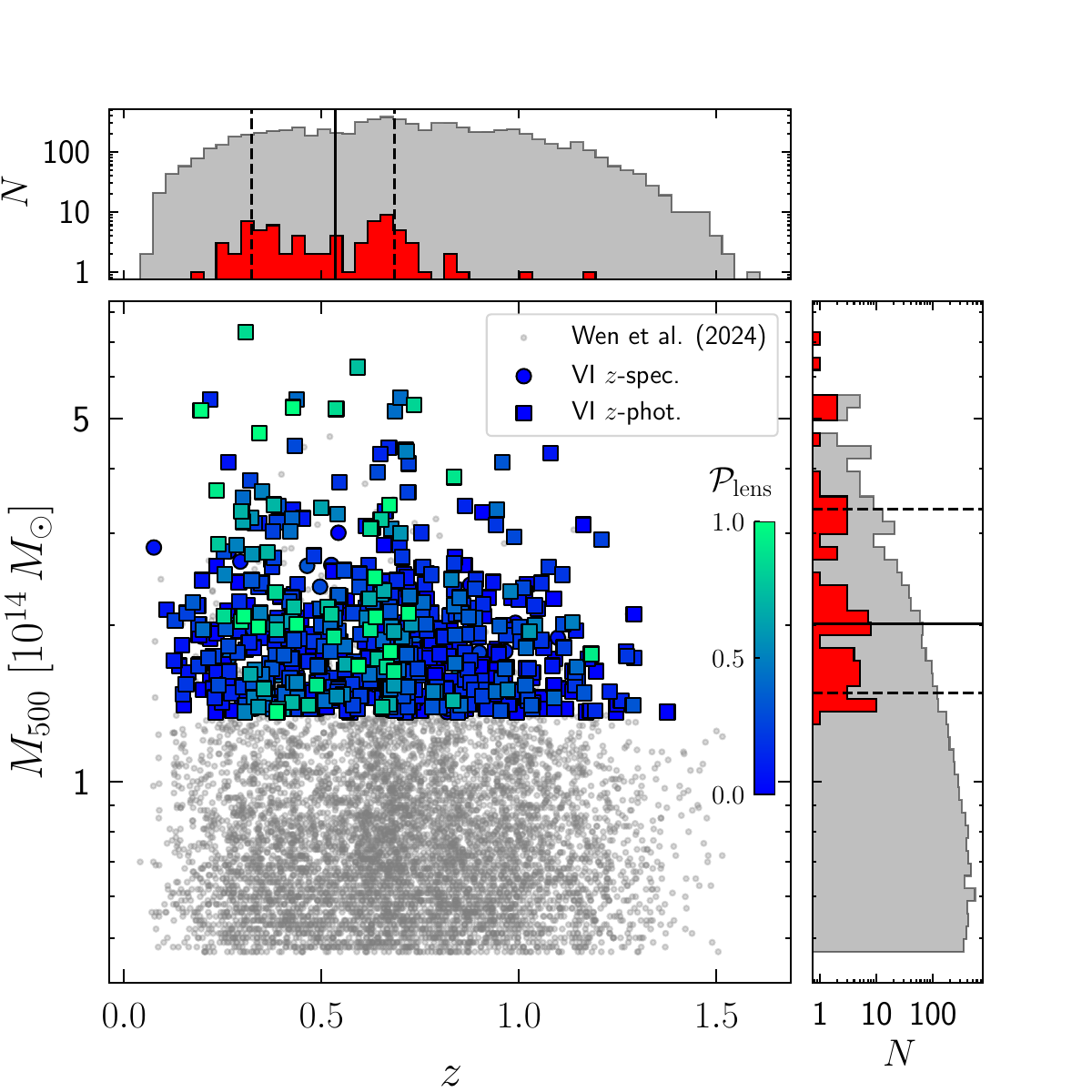}
\caption{Total mass-redshift distribution of the clusters visually inspected during the Phase 2 run. Small grey dots and grey histograms refer to the galaxy clusters detected in the DESI Legacy Imaging Surveys by \cite{Wen_2024} and located within the Q1 $63.1\,\mathrm{deg^2}$ area. The squares and dots correspond, respectively, to the galaxy clusters with measured photometric or spectroscopic redshifts, colour-coded according to the lensing probability, $\mathcal{P}_{\mathrm{lens}}$. The total mass and redshift distributions of the galaxy clusters with $\mathcal{P}_{\mathrm{lens}}>0.5$ are shown as red histograms, while the black lines represent the 16th (dashed line), 50th (solid line), and 84th (dashed line) percentiles of these distributions. The effect of the selection criterion ($\lambda_{500}>30$) adopted to select the galaxy clusters for the visual inspection is clearly visible.}
\label{fig:massz}
\end{figure}

Although a detailed study of the physical properties of the newly observed galaxy cluster strong gravitational lenses is beyond the scope of this work, we present some preliminary results from our analysis in this section.
In \Fig\ref{fig:massz}, we plot the distribution of $M_{500}$ cluster masses as a function of redshift. The markers representing the visually inspected clusters are colour-coded according to their $\mathcal{P}_{\mathrm{lens}}$ values. The figure reveals that most galaxy cluster lenses are in the redshift range $0.32<z<0.68$. This result can be attributed to a combination of two main factors. On the one hand, the wavelength coverage of the \Euclid\ photometric filters, the EWS image depth, and the higher resolution of the \IE\ band compared to the near-IR bands are designed to facilitate the identification of most of the galaxies at $z<2$. On the other hand, as shown by \citet[][see also \citealt{Meneghetti_2023}]{Meneghetti_2013} using cosmological simulations, the lensing cross-section for giant arcs for sources at that redshift is maximised for cluster lenses at $z\simeq0.3$. 

\begin{figure}[t]
\centering
\includegraphics[width=\linewidth]{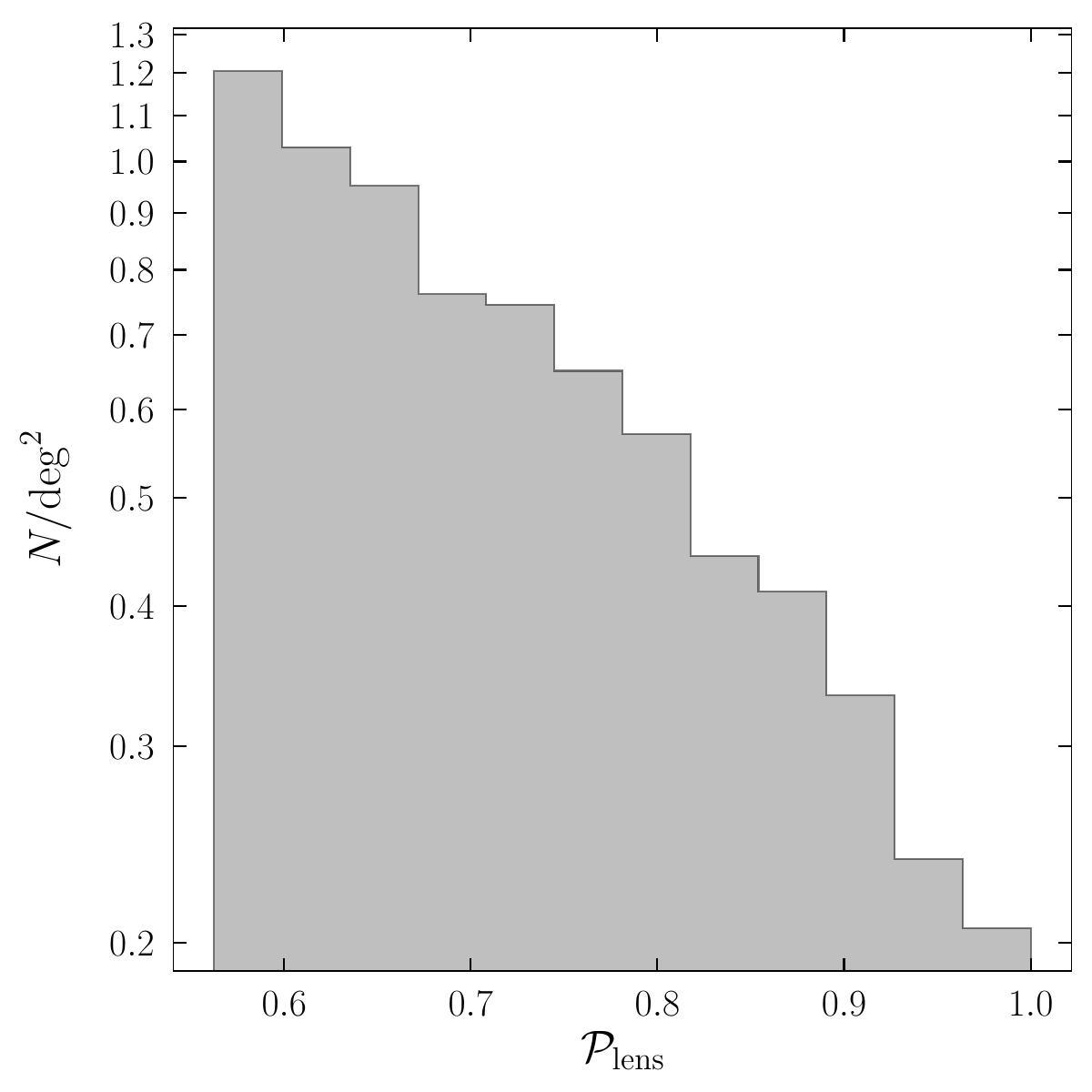}
\caption{Inverse cumulative distribution of the galaxy cluster number density as a function of the lensing probability, $\mathcal{P}_{\mathrm{lens}}$, assigned during the second phase of the visual inspection process.}
\label{fig:Ndeg}
\end{figure}

Based on our census of galaxy cluster strong gravitational lenses identified in the $63.1\,\mathrm{deg}^2$ covered by EDF-N, EDF-S, and EDF-F, we can provide an immediate estimate of the expected number density of cluster lenses per square degree detectable in the EWS. In \Fig\ref{fig:Ndeg}, we show the inverse cumulative distribution of the number density of cluster lenses as a function of the lensing probability. Specifically, we expect approximately $0.3\,\mathrm{deg}^{-2}$  cluster lenses with $\mathcal{P}_{\mathrm{lens}}>0.9$. By lowering the lensing probability threshold to $\mathcal{P}_{\mathrm{lens}}>0.5$, the expected number density of cluster lenses increases to $1.2\,\mathrm{deg}^{-2}$. Given the final EWS sky coverage of approximately 14\,000\,$\mathrm{deg}^2$, we predict that \Euclid\ will detect more than $4200$ cluster lenses exhibiting unequivocal evidence of extended, bright giant arcs and multiple images. It is important to note that these estimates are likely conservative, since they do not account for newly discovered cluster lenses that are not included in existing catalogues. 

In \Fig\ref{fig:bullet}, we show as an example the rectangles drawn by the inspectors in the field of ACT-CLJ0411.2$-$4819, highlighting the strong lensing features they identified. The red pixels, corresponding to regions where many rectangles overlap, indicate areas of the clusters that contain prominent giant arcs and multiple images. These rectangles will provide the sample to train deep-learning models to find similar features automatically in future \Euclid data. 

Among the secure cluster lenses detected in Q1, the cluster identified as ACT-CLJ0411.2$-$4819, shown in \Fig\ref{fig:bullet}, deserves special mention. This is a merging cluster that exhibits a variety of bright and extended strong lensing features, including tangential and radial giant arcs at different redshifts. Additionally, several smaller galaxy groups surrounding the main cluster, each displaying evident strong lensing features, make this cluster an excellent candidate for future follow-up observations.

Finally, we also identified several galaxy-galaxy strong lensing events in cluster fields. This is expected since the dense galaxy cluster environments enhance the strong lensing cross section of galaxies in these fields \citep[e.g.,][]{Meneghetti_2020}. 

\section{Conclusions}
\label{sec:conclusions}
We have presented the first catalogue of strong lensing galaxy clusters observed by  \Euclid. These clusters were part of the catalogue of optically-selected galaxy clusters by \cite{Wen_2024} supplemented with other clusters. Most of them were observed from space for the first time, thanks to \Euclid.  For this reason, only a few of these galaxy clusters were known to host some strong lensing features. 
Thanks to the spatial resolution and depth of \Euclid observations, we identified several gravitational arcs and multiple images of galaxies in the cluster backgrounds. 

To find these features and classify clusters as lenses and non-lenses, we visually inspected colour images created by combining the \Euclid observations in the \IE, \YE, and \HE bands. The search was blindly carried out by $44$ inspectors using a new interactive visualisation tool. We found 83 strong gravitational lenses among over a thousand  inspected clusters, 14 of which exhibit secure strong lensing features such as tangential and radial arcs and multiple images. To our knowledge, only two arcs among those we have found had been previously identified in ground-based observations \citep[the brightest arcs in J174330.4$+$634142, a.k.a. A2280, and ACT-CLJ0411.2$-$4819][]{Gioia_1995,Bayliss_2016}. Only three of the identified gravitational lenses, marked with stars ($\star$) in Tables\,\ref{table:phase1_lenses} and \ref{table:phase2_lenses}, had been previously observed by the {\it Hubble} Space Telescope, with one of them, marked with a dagger ($\dagger$), also observed by the {\it James Webb} Space Telescope. For the remaining 80 lenses, \Euclid\ has provided the first space-based optical observations and the spatial resolution necessary to discover many more strong lensing features. Considering the number of newly observed lensing galaxy clusters and the high quality of \Euclid\ imaging data, even at the depth of the EWS, \Euclid\ will increase the sample of known gravitational lenses by more than one order of magnitude, paving the way for an unprecedented advancement in the field of gravitational lensing by galaxy clusters.

In the sample we have inspected, we found that around $0.3$ clusters per square degree contain evident strong lensing features. Based on this number density, we expect that \Euclid\ will observe more than $4200$ strong lensing clusters in the EWS, assuming optical selection above a minimum richness of 30. This estimate is consistent with the forecasts of \cite{Boldrin_2016}, based on simulations in the $\Lambda$CDM cosmological model, although the impact of the selection function needs to be studied in more detail. 

The redshift distribution of the strong lensing clusters in our sample peaks at $z\simeq 0.5$, also consistent with expectations \citep[e.g.,][]{Boldrin_2012, Meneghetti_2013, Boldrin_2016}. Based on the galaxy distributions, the sample does not appear to be biased in terms of cluster dynamical state. Few lenses show elongated or multimodal spatial distributions of galaxies that are likely cluster members. Other lenses have more axially symmetric distributions of cluster galaxies.

Soon, we expect to find more strong lensing clusters in the upcoming  \Euclid\ data. These observations will enable many new studies based on strong gravitational lensing, aimed at understanding the nature of dark matter and dark energy, constraining cosmological parameters, studying high-redshift sources, and more. For many of these lensing applications, combining the \Euclid data with complementary observations, including spectroscopy and observations in the X-ray, submillimetre, and radio domains will be crucial. Exploiting the synergy between \Euclid\ and optical time-domain surveys, in particular the Vera C. Rubin Observatory’s Legacy Survey of Space and Time \citep{Guy_2022}, will also enable discovery of explosive transients that are gravitationally lensed by the strong lensing galaxy clusters.

\Euclid is accumulating data at around $6.25\,\deg^2$ per day. The fast data volume growth makes our visual inspection approach inefficient in searching for strong lenses. Given that the Phase 2 run took approximately three weeks to be completed, we estimate that visually inspecting the entire EWS area, 260 times larger than the Q1 area, would require more than 15 years. As seen in searches for galaxy-galaxy strong lenses, citizen science may help mitigate this problem \citep{Q1-SP048, Q1-SP059}. However, despite being less numerous, galaxy clusters are much more complex lenses than galaxies. They produce a variety of image configurations that inspectors cannot easily recognise without proper training. For this reason, automating the search for strong lensing features in galaxy clusters is a priority. In this work, we identified several tens of gravitational arcs and arclets. We will combine them with simulated data to train algorithms based on deep learning, such as different architectures of convolutional neural networks, to find features like gravitational arcs, arclets, and multiple images. 

Our work demonstrates the huge potential of the \Euclid mission for discovering new strong lensing clusters. We identified several lenses with multiple gravitational arcs, implying large strong lensing cross-sections. These galaxy clusters will be re-observed by \Euclid multiple times during the course of the mission as part of the EDF survey. These future observations will allow us to find additional arcs and multiple images by reaching fainter magnitude limits. Thus, we will collect many more observables to build robust total mass models of these lenses. In addition, by comparing observations taken at different times, we may be able to detect lensed transients. The observations of such sources would enable further science cases. For example, in the case of multiply imaged time-variable sources, it may be possible to measure the cosmic expansion rate \citep{Kelly_2015,Kelly_2016,Grillo_2015,Grillo_2018,Grillo2024}.   

\begin{acknowledgements}
 We acknowledge financial support through grants PRIN-MIUR 2017WSCC32 and 2020SKSTHZ. MM acknowledges support from the Italian Space Agency (ASI) through contract ``Euclid - Phase E''. AA acknowledges financial support through the project PID2022-138896NB-C51 (MCIU/AEI/MINECO/FEDER, UE) Ministerio de Ciencia, Investigaci\'on y Universidades. MJ is supported by the United Kingdom Research and Innovation (UKRI) Future Leaders Fellowship ``Using Cosmic Beasts to uncover the Nature of Dark Matter'' (grant number MR/S017216/1 and MR/X006069/1).
 \AckERO \AckQone
  \AckEC \AckCosmoHub\ Based on data from UNIONS, a scientific collaboration using
three Hawaii-based telescopes: CFHT, Pan-STARRS, and Subaru
\url{www.skysurvey.cc}\,. Based on
data from the Dark Energy Camera (DECam) on the Blanco 4-m Telescope
at CTIO in Chile \url{https://www.darkenergysurvey.org}\,. This work uses results from the ESA mission {\it Gaia},
whose data are being processed by the Gaia Data Processing and
Analysis Consortium \url{https://www.cosmos.esa.int/gaia}\,.
\end{acknowledgements}

%
%

\bibliographystyle{aa}
\bibliography{bibliography, Euclid_new, Q1}

\label{LastPage}

\end{document}

%% file: authorlist.tex
\newcommand{\orcid}[1]{} 
\author{Euclid Collaboration: P.~Bergamini\orcid{0000-0003-1383-9414}\thanks{\email{pietro.bergamini@inaf.it}}\inst{\ref{aff1},\ref{aff2}}
\and M.~Meneghetti\orcid{0000-0003-1225-7084}\inst{\ref{aff2},\ref{aff3}}
\and A.~Acebron\orcid{0000-0003-3108-9039}\inst{\ref{aff4}}
\and B.~Cl\'ement\orcid{0000-0002-7966-3661}\inst{\ref{aff5},\ref{aff6}}
\and M.~Bolzonella\orcid{0000-0003-3278-4607}\inst{\ref{aff2}}
\and C.~Grillo\orcid{0000-0002-5926-7143}\inst{\ref{aff1},\ref{aff7}}
\and P.~Rosati\orcid{0000-0002-6813-0632}\inst{\ref{aff8},\ref{aff2}}
\and D.~Abriola\orcid{0009-0005-4230-3266}\inst{\ref{aff1}}
\and J.~A.~Acevedo~Barroso\orcid{0000-0002-9654-1711}\inst{\ref{aff5}}
\and G.~Angora\orcid{0000-0002-0316-6562}\inst{\ref{aff9},\ref{aff8}}
\and L.~Bazzanini\orcid{0000-0003-0727-0137}\inst{\ref{aff8},\ref{aff2}}
\and R.~Cabanac\orcid{0000-0001-6679-2600}\inst{\ref{aff10}}
\and B.~C.~Nagam\orcid{0000-0002-3724-7694}\inst{\ref{aff11},\ref{aff12}}
\and A.~R.~Cooray\orcid{0000-0002-3892-0190}\inst{\ref{aff13}}
\and G.~Despali\orcid{0000-0001-6150-4112}\inst{\ref{aff14},\ref{aff2},\ref{aff3}}
\and G.~Di~Rosa\orcid{0009-0001-9416-0923}\inst{\ref{aff8}}
\and J.~M.~Diego\orcid{0000-0001-9065-3926}\inst{\ref{aff4}}
\and M.~Fogliardi\orcid{0009-0006-4964-5311}\inst{\ref{aff8}}
\and A.~Galan\orcid{0000-0003-2547-9815}\inst{\ref{aff15},\ref{aff16}}
\and R.~Gavazzi\orcid{0000-0002-5540-6935}\inst{\ref{aff17},\ref{aff18}}
\and G.~Granata\orcid{0000-0002-9512-3788}\inst{\ref{aff19}}
\and N.~B.~Hogg\orcid{0000-0001-9346-4477}\inst{\ref{aff20}}
\and K.~Jahnke\orcid{0000-0003-3804-2137}\inst{\ref{aff21}}
\and L.~Leuzzi\orcid{0009-0006-4479-7017}\inst{\ref{aff14},\ref{aff2}}
\and T.~Li\orcid{0009-0005-5008-0381}\inst{\ref{aff19}}
\and M.~Lombardi\orcid{0000-0002-3336-4965}\inst{\ref{aff1}}
\and G.~Mahler\orcid{0000-0003-3266-2001}\inst{\ref{aff22},\ref{aff23},\ref{aff24}}
\and A.~Manj\'on-Garc\'ia\orcid{0000-0002-7413-8825}\inst{\ref{aff25}}
\and R.~B.~Metcalf\orcid{0000-0003-3167-2574}\inst{\ref{aff14},\ref{aff2}}
\and M.~Oguri\orcid{0000-0003-3484-399X}\inst{\ref{aff26},\ref{aff27}}
\and C.~Olave\inst{\ref{aff28}}
\and J.~M.~Palencia\orcid{0000-0003-0942-817X}\inst{\ref{aff4}}
\and J.~Richard\orcid{0000-0001-5492-1049}\inst{\ref{aff29}}
\and K.~Rojas\orcid{0000-0003-1391-6854}\inst{\ref{aff30},\ref{aff19}}
\and L.~R.~Ecker\orcid{0009-0005-3508-2469}\inst{\ref{aff31},\ref{aff32}}
\and C.~Scarlata\orcid{0000-0002-9136-8876}\inst{\ref{aff11}}
\and M.~Schirmer\orcid{0000-0003-2568-9994}\inst{\ref{aff21}}
\and S.~Schuldt\orcid{0000-0003-2497-6334}\inst{\ref{aff1},\ref{aff7}}
\and D.~Sluse\orcid{0000-0001-6116-2095}\inst{\ref{aff22}}
\and G.~P.~Smith\orcid{0000-0003-4494-8277}\inst{\ref{aff33},\ref{aff34}}
\and C.~Tortora\orcid{0000-0001-7958-6531}\inst{\ref{aff9}}
\and G.~Vernardos\orcid{0000-0001-8554-7248}\inst{\ref{aff35},\ref{aff36}}
\and G.~L.~Walth\orcid{0000-0002-6313-6808}\inst{\ref{aff37}}
\and J.~Wilde\orcid{0000-0002-4460-7379}\inst{\ref{aff38}}
\and Y.~Xie\inst{\ref{aff5},\ref{aff39}}
\and M.~Zumalacarregui\orcid{0000-0002-9943-6490}\inst{\ref{aff40}}
\and N.~Aghanim\orcid{0000-0002-6688-8992}\inst{\ref{aff41}}
\and B.~Altieri\orcid{0000-0003-3936-0284}\inst{\ref{aff42}}
\and A.~Amara\inst{\ref{aff43}}
\and L.~Amendola\orcid{0000-0002-0835-233X}\inst{\ref{aff44}}
\and S.~Andreon\orcid{0000-0002-2041-8784}\inst{\ref{aff45}}
\and N.~Auricchio\orcid{0000-0003-4444-8651}\inst{\ref{aff2}}
\and H.~Aussel\orcid{0000-0002-1371-5705}\inst{\ref{aff46}}
\and C.~Baccigalupi\orcid{0000-0002-8211-1630}\inst{\ref{aff47},\ref{aff48},\ref{aff49},\ref{aff50}}
\and M.~Baldi\orcid{0000-0003-4145-1943}\inst{\ref{aff28},\ref{aff2},\ref{aff3}}
\and A.~Balestra\orcid{0000-0002-6967-261X}\inst{\ref{aff51}}
\and S.~Bardelli\orcid{0000-0002-8900-0298}\inst{\ref{aff2}}
\and A.~Basset\inst{\ref{aff52}}
\and P.~Battaglia\orcid{0000-0002-7337-5909}\inst{\ref{aff2}}
\and R.~Bender\orcid{0000-0001-7179-0626}\inst{\ref{aff32},\ref{aff31}}
\and A.~Biviano\orcid{0000-0002-0857-0732}\inst{\ref{aff48},\ref{aff47}}
\and A.~Bonchi\orcid{0000-0002-2667-5482}\inst{\ref{aff53}}
\and D.~Bonino\orcid{0000-0002-3336-9977}\inst{\ref{aff54}}
\and E.~Branchini\orcid{0000-0002-0808-6908}\inst{\ref{aff55},\ref{aff56},\ref{aff45}}
\and M.~Brescia\orcid{0000-0001-9506-5680}\inst{\ref{aff57},\ref{aff9}}
\and J.~Brinchmann\orcid{0000-0003-4359-8797}\inst{\ref{aff58},\ref{aff59}}
\and A.~Caillat\inst{\ref{aff17}}
\and S.~Camera\orcid{0000-0003-3399-3574}\inst{\ref{aff60},\ref{aff61},\ref{aff54}}
\and G.~Ca\~nas-Herrera\orcid{0000-0003-2796-2149}\inst{\ref{aff62},\ref{aff63},\ref{aff64}}
\and V.~Capobianco\orcid{0000-0002-3309-7692}\inst{\ref{aff54}}
\and C.~Carbone\orcid{0000-0003-0125-3563}\inst{\ref{aff7}}
\and J.~Carretero\orcid{0000-0002-3130-0204}\inst{\ref{aff65},\ref{aff66}}
\and S.~Casas\orcid{0000-0002-4751-5138}\inst{\ref{aff67}}
\and F.~J.~Castander\orcid{0000-0001-7316-4573}\inst{\ref{aff68},\ref{aff69}}
\and M.~Castellano\orcid{0000-0001-9875-8263}\inst{\ref{aff70}}
\and G.~Castignani\orcid{0000-0001-6831-0687}\inst{\ref{aff2}}
\and S.~Cavuoti\orcid{0000-0002-3787-4196}\inst{\ref{aff9},\ref{aff71}}
\and K.~C.~Chambers\orcid{0000-0001-6965-7789}\inst{\ref{aff72}}
\and A.~Cimatti\inst{\ref{aff73}}
\and C.~Colodro-Conde\inst{\ref{aff74}}
\and G.~Congedo\orcid{0000-0003-2508-0046}\inst{\ref{aff75}}
\and C.~J.~Conselice\orcid{0000-0003-1949-7638}\inst{\ref{aff76}}
\and L.~Conversi\orcid{0000-0002-6710-8476}\inst{\ref{aff77},\ref{aff42}}
\and Y.~Copin\orcid{0000-0002-5317-7518}\inst{\ref{aff78}}
\and F.~Courbin\orcid{0000-0003-0758-6510}\inst{\ref{aff38},\ref{aff79}}
\and H.~M.~Courtois\orcid{0000-0003-0509-1776}\inst{\ref{aff80}}
\and M.~Cropper\orcid{0000-0003-4571-9468}\inst{\ref{aff81}}
\and A.~Da~Silva\orcid{0000-0002-6385-1609}\inst{\ref{aff82},\ref{aff83}}
\and H.~Degaudenzi\orcid{0000-0002-5887-6799}\inst{\ref{aff84}}
\and G.~De~Lucia\orcid{0000-0002-6220-9104}\inst{\ref{aff48}}
\and A.~M.~Di~Giorgio\orcid{0000-0002-4767-2360}\inst{\ref{aff85}}
\and C.~Dolding\orcid{0009-0003-7199-6108}\inst{\ref{aff81}}
\and H.~Dole\orcid{0000-0002-9767-3839}\inst{\ref{aff41}}
\and F.~Dubath\orcid{0000-0002-6533-2810}\inst{\ref{aff84}}
\and X.~Dupac\inst{\ref{aff42}}
\and S.~Dusini\orcid{0000-0002-1128-0664}\inst{\ref{aff86}}
\and A.~Ealet\orcid{0000-0003-3070-014X}\inst{\ref{aff78}}
\and S.~Escoffier\orcid{0000-0002-2847-7498}\inst{\ref{aff87}}
\and M.~Fabricius\orcid{0000-0002-7025-6058}\inst{\ref{aff32},\ref{aff31}}
\and M.~Farina\orcid{0000-0002-3089-7846}\inst{\ref{aff85}}
\and R.~Farinelli\inst{\ref{aff2}}
\and F.~Faustini\orcid{0000-0001-6274-5145}\inst{\ref{aff70},\ref{aff53}}
\and S.~Ferriol\inst{\ref{aff78}}
\and F.~Finelli\orcid{0000-0002-6694-3269}\inst{\ref{aff2},\ref{aff88}}
\and P.~Fosalba\orcid{0000-0002-1510-5214}\inst{\ref{aff69},\ref{aff68}}
\and S.~Fotopoulou\orcid{0000-0002-9686-254X}\inst{\ref{aff89}}
\and M.~Frailis\orcid{0000-0002-7400-2135}\inst{\ref{aff48}}
\and E.~Franceschi\orcid{0000-0002-0585-6591}\inst{\ref{aff2}}
\and M.~Fumana\orcid{0000-0001-6787-5950}\inst{\ref{aff7}}
\and S.~Galeotta\orcid{0000-0002-3748-5115}\inst{\ref{aff48}}
\and K.~George\orcid{0000-0002-1734-8455}\inst{\ref{aff31}}
\and B.~Gillis\orcid{0000-0002-4478-1270}\inst{\ref{aff75}}
\and C.~Giocoli\orcid{0000-0002-9590-7961}\inst{\ref{aff2},\ref{aff3}}
\and P.~G\'omez-Alvarez\orcid{0000-0002-8594-5358}\inst{\ref{aff90},\ref{aff42}}
\and J.~Gracia-Carpio\inst{\ref{aff32}}
\and B.~R.~Granett\orcid{0000-0003-2694-9284}\inst{\ref{aff45}}
\and A.~Grazian\orcid{0000-0002-5688-0663}\inst{\ref{aff51}}
\and F.~Grupp\inst{\ref{aff32},\ref{aff31}}
\and L.~Guzzo\orcid{0000-0001-8264-5192}\inst{\ref{aff1},\ref{aff45},\ref{aff91}}
\and S.~V.~H.~Haugan\orcid{0000-0001-9648-7260}\inst{\ref{aff92}}
\and H.~Hoekstra\orcid{0000-0002-0641-3231}\inst{\ref{aff64}}
\and W.~Holmes\inst{\ref{aff93}}
\and F.~Hormuth\inst{\ref{aff94}}
\and A.~Hornstrup\orcid{0000-0002-3363-0936}\inst{\ref{aff95},\ref{aff96}}
\and P.~Hudelot\inst{\ref{aff18}}
\and M.~Jhabvala\inst{\ref{aff97}}
\and B.~Joachimi\orcid{0000-0001-7494-1303}\inst{\ref{aff98}}
\and E.~Keih\"anen\orcid{0000-0003-1804-7715}\inst{\ref{aff99}}
\and S.~Kermiche\orcid{0000-0002-0302-5735}\inst{\ref{aff87}}
\and A.~Kiessling\orcid{0000-0002-2590-1273}\inst{\ref{aff93}}
\and M.~Kilbinger\orcid{0000-0001-9513-7138}\inst{\ref{aff46}}
\and R.~Kohley\inst{\ref{aff42}}
\and B.~Kubik\orcid{0009-0006-5823-4880}\inst{\ref{aff78}}
\and K.~Kuijken\orcid{0000-0002-3827-0175}\inst{\ref{aff64}}
\and M.~K\"ummel\orcid{0000-0003-2791-2117}\inst{\ref{aff31}}
\and M.~Kunz\orcid{0000-0002-3052-7394}\inst{\ref{aff100}}
\and H.~Kurki-Suonio\orcid{0000-0002-4618-3063}\inst{\ref{aff101},\ref{aff102}}
\and O.~Lahav\orcid{0000-0002-1134-9035}\inst{\ref{aff98}}
\and R.~Laureijs\inst{\ref{aff12}}
\and Q.~Le~Boulc'h\inst{\ref{aff103}}
\and A.~M.~C.~Le~Brun\orcid{0000-0002-0936-4594}\inst{\ref{aff104}}
\and D.~Le~Mignant\orcid{0000-0002-5339-5515}\inst{\ref{aff17}}
\and P.~Liebing\inst{\ref{aff81}}
\and S.~Ligori\orcid{0000-0003-4172-4606}\inst{\ref{aff54}}
\and P.~B.~Lilje\orcid{0000-0003-4324-7794}\inst{\ref{aff92}}
\and V.~Lindholm\orcid{0000-0003-2317-5471}\inst{\ref{aff101},\ref{aff102}}
\and I.~Lloro\orcid{0000-0001-5966-1434}\inst{\ref{aff105}}
\and G.~Mainetti\orcid{0000-0003-2384-2377}\inst{\ref{aff103}}
\and D.~Maino\inst{\ref{aff1},\ref{aff7},\ref{aff91}}
\and E.~Maiorano\orcid{0000-0003-2593-4355}\inst{\ref{aff2}}
\and O.~Mansutti\orcid{0000-0001-5758-4658}\inst{\ref{aff48}}
\and S.~Marcin\inst{\ref{aff106}}
\and O.~Marggraf\orcid{0000-0001-7242-3852}\inst{\ref{aff107}}
\and M.~Martinelli\orcid{0000-0002-6943-7732}\inst{\ref{aff70},\ref{aff108}}
\and N.~Martinet\orcid{0000-0003-2786-7790}\inst{\ref{aff17}}
\and F.~Marulli\orcid{0000-0002-8850-0303}\inst{\ref{aff14},\ref{aff2},\ref{aff3}}
\and R.~Massey\orcid{0000-0002-6085-3780}\inst{\ref{aff24}}
\and S.~Maurogordato\inst{\ref{aff109}}
\and E.~Medinaceli\orcid{0000-0002-4040-7783}\inst{\ref{aff2}}
\and S.~Mei\orcid{0000-0002-2849-559X}\inst{\ref{aff110},\ref{aff111}}
\and M.~Melchior\inst{\ref{aff30}}
\and Y.~Mellier\inst{\ref{aff112},\ref{aff18}}
\and E.~Merlin\orcid{0000-0001-6870-8900}\inst{\ref{aff70}}
\and G.~Meylan\inst{\ref{aff5}}
\and A.~Mora\orcid{0000-0002-1922-8529}\inst{\ref{aff113}}
\and M.~Moresco\orcid{0000-0002-7616-7136}\inst{\ref{aff14},\ref{aff2}}
\and L.~Moscardini\orcid{0000-0002-3473-6716}\inst{\ref{aff14},\ref{aff2},\ref{aff3}}
\and S.~Mourre\orcid{0009-0005-9047-0691}\inst{\ref{aff109},\ref{aff114}}
\and R.~Nakajima\orcid{0009-0009-1213-7040}\inst{\ref{aff107}}
\and C.~Neissner\orcid{0000-0001-8524-4968}\inst{\ref{aff115},\ref{aff66}}
\and R.~C.~Nichol\orcid{0000-0003-0939-6518}\inst{\ref{aff43}}
\and S.-M.~Niemi\inst{\ref{aff62}}
\and J.~W.~Nightingale\orcid{0000-0002-8987-7401}\inst{\ref{aff116}}
\and C.~Padilla\orcid{0000-0001-7951-0166}\inst{\ref{aff115}}
\and S.~Paltani\orcid{0000-0002-8108-9179}\inst{\ref{aff84}}
\and F.~Pasian\orcid{0000-0002-4869-3227}\inst{\ref{aff48}}
\and K.~Pedersen\inst{\ref{aff117}}
\and W.~J.~Percival\orcid{0000-0002-0644-5727}\inst{\ref{aff118},\ref{aff119},\ref{aff120}}
\and V.~Pettorino\inst{\ref{aff62}}
\and S.~Pires\orcid{0000-0002-0249-2104}\inst{\ref{aff46}}
\and G.~Polenta\orcid{0000-0003-4067-9196}\inst{\ref{aff53}}
\and M.~Poncet\inst{\ref{aff52}}
\and L.~A.~Popa\inst{\ref{aff121}}
\and L.~Pozzetti\orcid{0000-0001-7085-0412}\inst{\ref{aff2}}
\and F.~Raison\orcid{0000-0002-7819-6918}\inst{\ref{aff32}}
\and R.~Rebolo\orcid{0000-0003-3767-7085}\inst{\ref{aff74},\ref{aff122},\ref{aff123}}
\and A.~Renzi\orcid{0000-0001-9856-1970}\inst{\ref{aff124},\ref{aff86}}
\and J.~Rhodes\orcid{0000-0002-4485-8549}\inst{\ref{aff93}}
\and G.~Riccio\inst{\ref{aff9}}
\and E.~Romelli\orcid{0000-0003-3069-9222}\inst{\ref{aff48}}
\and M.~Roncarelli\orcid{0000-0001-9587-7822}\inst{\ref{aff2}}
\and B.~Rusholme\orcid{0000-0001-7648-4142}\inst{\ref{aff37}}
\and R.~Saglia\orcid{0000-0003-0378-7032}\inst{\ref{aff31},\ref{aff32}}
\and Z.~Sakr\orcid{0000-0002-4823-3757}\inst{\ref{aff44},\ref{aff10},\ref{aff125}}
\and D.~Sapone\orcid{0000-0001-7089-4503}\inst{\ref{aff126}}
\and B.~Sartoris\orcid{0000-0003-1337-5269}\inst{\ref{aff31},\ref{aff48}}
\and J.~A.~Schewtschenko\orcid{0000-0002-4913-6393}\inst{\ref{aff75}}
\and P.~Schneider\orcid{0000-0001-8561-2679}\inst{\ref{aff107}}
\and A.~Secroun\orcid{0000-0003-0505-3710}\inst{\ref{aff87}}
\and G.~Seidel\orcid{0000-0003-2907-353X}\inst{\ref{aff21}}
\and M.~Seiffert\orcid{0000-0002-7536-9393}\inst{\ref{aff93}}
\and S.~Serrano\orcid{0000-0002-0211-2861}\inst{\ref{aff69},\ref{aff127},\ref{aff68}}
\and P.~Simon\inst{\ref{aff107}}
\and C.~Sirignano\orcid{0000-0002-0995-7146}\inst{\ref{aff124},\ref{aff86}}
\and G.~Sirri\orcid{0000-0003-2626-2853}\inst{\ref{aff3}}
\and A.~Spurio~Mancini\orcid{0000-0001-5698-0990}\inst{\ref{aff128}}
\and L.~Stanco\orcid{0000-0002-9706-5104}\inst{\ref{aff86}}
\and J.~Steinwagner\orcid{0000-0001-7443-1047}\inst{\ref{aff32}}
\and P.~Tallada-Cresp\'{i}\orcid{0000-0002-1336-8328}\inst{\ref{aff65},\ref{aff66}}
\and A.~N.~Taylor\inst{\ref{aff75}}
\and H.~I.~Teplitz\orcid{0000-0002-7064-5424}\inst{\ref{aff129}}
\and I.~Tereno\inst{\ref{aff82},\ref{aff130}}
\and N.~Tessore\orcid{0000-0002-9696-7931}\inst{\ref{aff98}}
\and S.~Toft\orcid{0000-0003-3631-7176}\inst{\ref{aff131},\ref{aff132}}
\and R.~Toledo-Moreo\orcid{0000-0002-2997-4859}\inst{\ref{aff133}}
\and F.~Torradeflot\orcid{0000-0003-1160-1517}\inst{\ref{aff66},\ref{aff65}}
\and A.~Tsyganov\inst{\ref{aff134}}
\and I.~Tutusaus\orcid{0000-0002-3199-0399}\inst{\ref{aff10}}
\and E.~A.~Valentijn\inst{\ref{aff12}}
\and L.~Valenziano\orcid{0000-0002-1170-0104}\inst{\ref{aff2},\ref{aff88}}
\and J.~Valiviita\orcid{0000-0001-6225-3693}\inst{\ref{aff101},\ref{aff102}}
\and T.~Vassallo\orcid{0000-0001-6512-6358}\inst{\ref{aff31},\ref{aff48}}
\and G.~Verdoes~Kleijn\orcid{0000-0001-5803-2580}\inst{\ref{aff12}}
\and A.~Veropalumbo\orcid{0000-0003-2387-1194}\inst{\ref{aff45},\ref{aff56},\ref{aff55}}
\and Y.~Wang\orcid{0000-0002-4749-2984}\inst{\ref{aff129}}
\and J.~Weller\orcid{0000-0002-8282-2010}\inst{\ref{aff31},\ref{aff32}}
\and A.~Zacchei\orcid{0000-0003-0396-1192}\inst{\ref{aff48},\ref{aff47}}
\and G.~Zamorani\orcid{0000-0002-2318-301X}\inst{\ref{aff2}}
\and F.~M.~Zerbi\inst{\ref{aff45}}
\and E.~Zucca\orcid{0000-0002-5845-8132}\inst{\ref{aff2}}
\and V.~Allevato\orcid{0000-0001-7232-5152}\inst{\ref{aff9}}
\and M.~Ballardini\orcid{0000-0003-4481-3559}\inst{\ref{aff8},\ref{aff135},\ref{aff2}}
\and E.~Bozzo\orcid{0000-0002-8201-1525}\inst{\ref{aff84}}
\and C.~Burigana\orcid{0000-0002-3005-5796}\inst{\ref{aff136},\ref{aff88}}
\and A.~Cappi\inst{\ref{aff2},\ref{aff109}}
\and P.~Casenove\orcid{0009-0006-6736-1670}\inst{\ref{aff52}}
\and D.~Di~Ferdinando\inst{\ref{aff3}}
\and J.~A.~Escartin~Vigo\inst{\ref{aff32}}
\and L.~Gabarra\orcid{0000-0002-8486-8856}\inst{\ref{aff137}}
\and J.~Mart\'{i}n-Fleitas\orcid{0000-0002-8594-569X}\inst{\ref{aff113}}
\and S.~Matthew\orcid{0000-0001-8448-1697}\inst{\ref{aff75}}
\and M.~Maturi\orcid{0000-0002-3517-2422}\inst{\ref{aff44},\ref{aff138}}
\and N.~Mauri\orcid{0000-0001-8196-1548}\inst{\ref{aff73},\ref{aff3}}
\and A.~A.~Nucita\inst{\ref{aff139},\ref{aff140},\ref{aff141}}
\and A.~Pezzotta\orcid{0000-0003-0726-2268}\inst{\ref{aff142},\ref{aff32}}
\and M.~P\"ontinen\orcid{0000-0001-5442-2530}\inst{\ref{aff101}}
\and C.~Porciani\orcid{0000-0002-7797-2508}\inst{\ref{aff107}}
\and I.~Risso\orcid{0000-0003-2525-7761}\inst{\ref{aff143}}
\and V.~Scottez\inst{\ref{aff112},\ref{aff144}}
\and M.~Sereno\orcid{0000-0003-0302-0325}\inst{\ref{aff2},\ref{aff3}}
\and M.~Tenti\orcid{0000-0002-4254-5901}\inst{\ref{aff3}}
\and M.~Viel\orcid{0000-0002-2642-5707}\inst{\ref{aff47},\ref{aff48},\ref{aff50},\ref{aff49},\ref{aff145}}
\and M.~Wiesmann\orcid{0009-0000-8199-5860}\inst{\ref{aff92}}
\and Y.~Akrami\orcid{0000-0002-2407-7956}\inst{\ref{aff146},\ref{aff147}}
\and I.~T.~Andika\orcid{0000-0001-6102-9526}\inst{\ref{aff15},\ref{aff16}}
\and S.~Anselmi\orcid{0000-0002-3579-9583}\inst{\ref{aff86},\ref{aff124},\ref{aff148}}
\and M.~Archidiacono\orcid{0000-0003-4952-9012}\inst{\ref{aff1},\ref{aff91}}
\and F.~Atrio-Barandela\orcid{0000-0002-2130-2513}\inst{\ref{aff149}}
\and C.~Benoist\inst{\ref{aff109}}
\and K.~Benson\inst{\ref{aff81}}
\and D.~Bertacca\orcid{0000-0002-2490-7139}\inst{\ref{aff124},\ref{aff51},\ref{aff86}}
\and M.~Bethermin\orcid{0000-0002-3915-2015}\inst{\ref{aff150}}
\and A.~Blanchard\orcid{0000-0001-8555-9003}\inst{\ref{aff10}}
\and L.~Blot\orcid{0000-0002-9622-7167}\inst{\ref{aff151},\ref{aff148}}
\and H.~B\"ohringer\orcid{0000-0001-8241-4204}\inst{\ref{aff32},\ref{aff152},\ref{aff153}}
\and S.~Borgani\orcid{0000-0001-6151-6439}\inst{\ref{aff154},\ref{aff47},\ref{aff48},\ref{aff49},\ref{aff145}}
\and M.~L.~Brown\orcid{0000-0002-0370-8077}\inst{\ref{aff76}}
\and S.~Bruton\orcid{0000-0002-6503-5218}\inst{\ref{aff155}}
\and A.~Calabro\orcid{0000-0003-2536-1614}\inst{\ref{aff70}}
\and B.~Camacho~Quevedo\orcid{0000-0002-8789-4232}\inst{\ref{aff69},\ref{aff68}}
\and F.~Caro\inst{\ref{aff70}}
\and C.~S.~Carvalho\inst{\ref{aff130}}
\and T.~Castro\orcid{0000-0002-6292-3228}\inst{\ref{aff48},\ref{aff49},\ref{aff47},\ref{aff145}}
\and F.~Cogato\orcid{0000-0003-4632-6113}\inst{\ref{aff14},\ref{aff2}}
\and O.~Cucciati\orcid{0000-0002-9336-7551}\inst{\ref{aff2}}
\and S.~Davini\orcid{0000-0003-3269-1718}\inst{\ref{aff56}}
\and F.~De~Paolis\orcid{0000-0001-6460-7563}\inst{\ref{aff139},\ref{aff140},\ref{aff141}}
\and G.~Desprez\orcid{0000-0001-8325-1742}\inst{\ref{aff12}}
\and A.~D\'iaz-S\'anchez\orcid{0000-0003-0748-4768}\inst{\ref{aff25}}
\and J.~J.~Diaz\inst{\ref{aff156}}
\and S.~Di~Domizio\orcid{0000-0003-2863-5895}\inst{\ref{aff55},\ref{aff56}}
\and P.-A.~Duc\orcid{0000-0003-3343-6284}\inst{\ref{aff150}}
\and A.~Enia\orcid{0000-0002-0200-2857}\inst{\ref{aff28},\ref{aff2}}
\and Y.~Fang\inst{\ref{aff31}}
\and A.~G.~Ferrari\orcid{0009-0005-5266-4110}\inst{\ref{aff3}}
\and P.~G.~Ferreira\orcid{0000-0002-3021-2851}\inst{\ref{aff137}}
\and A.~Finoguenov\orcid{0000-0002-4606-5403}\inst{\ref{aff101}}
\and A.~Fontana\orcid{0000-0003-3820-2823}\inst{\ref{aff70}}
\and A.~Franco\orcid{0000-0002-4761-366X}\inst{\ref{aff140},\ref{aff139},\ref{aff141}}
\and K.~Ganga\orcid{0000-0001-8159-8208}\inst{\ref{aff110}}
\and J.~Garc\'ia-Bellido\orcid{0000-0002-9370-8360}\inst{\ref{aff146}}
\and T.~Gasparetto\orcid{0000-0002-7913-4866}\inst{\ref{aff48}}
\and V.~Gautard\inst{\ref{aff157}}
\and E.~Gaztanaga\orcid{0000-0001-9632-0815}\inst{\ref{aff68},\ref{aff69},\ref{aff19}}
\and F.~Giacomini\orcid{0000-0002-3129-2814}\inst{\ref{aff3}}
\and F.~Gianotti\orcid{0000-0003-4666-119X}\inst{\ref{aff2}}
\and A.~H.~Gonzalez\orcid{0000-0002-0933-8601}\inst{\ref{aff158}}
\and G.~Gozaliasl\orcid{0000-0002-0236-919X}\inst{\ref{aff159},\ref{aff101}}
\and M.~Guidi\orcid{0000-0001-9408-1101}\inst{\ref{aff28},\ref{aff2}}
\and C.~M.~Gutierrez\orcid{0000-0001-7854-783X}\inst{\ref{aff160}}
\and A.~Hall\orcid{0000-0002-3139-8651}\inst{\ref{aff75}}
\and W.~G.~Hartley\inst{\ref{aff84}}
\and C.~Hern\'andez-Monteagudo\orcid{0000-0001-5471-9166}\inst{\ref{aff123},\ref{aff74}}
\and H.~Hildebrandt\orcid{0000-0002-9814-3338}\inst{\ref{aff161}}
\and J.~Hjorth\orcid{0000-0002-4571-2306}\inst{\ref{aff117}}
\and O.~Ilbert\orcid{0000-0002-7303-4397}\inst{\ref{aff17}}
\and M.~Jauzac\orcid{0000-0003-1974-8732}\inst{\ref{aff23},\ref{aff24},\ref{aff162},\ref{aff163}}
\and J.~J.~E.~Kajava\orcid{0000-0002-3010-8333}\inst{\ref{aff164},\ref{aff165}}
\and Y.~Kang\orcid{0009-0000-8588-7250}\inst{\ref{aff84}}
\and V.~Kansal\orcid{0000-0002-4008-6078}\inst{\ref{aff166},\ref{aff167}}
\and D.~Karagiannis\orcid{0000-0002-4927-0816}\inst{\ref{aff8},\ref{aff168}}
\and K.~Kiiveri\inst{\ref{aff99}}
\and C.~C.~Kirkpatrick\inst{\ref{aff99}}
\and S.~Kruk\orcid{0000-0001-8010-8879}\inst{\ref{aff42}}
\and J.~Le~Graet\orcid{0000-0001-6523-7971}\inst{\ref{aff87}}
\and L.~Legrand\orcid{0000-0003-0610-5252}\inst{\ref{aff169},\ref{aff170}}
\and M.~Lembo\orcid{0000-0002-5271-5070}\inst{\ref{aff8},\ref{aff135}}
\and F.~Lepori\orcid{0009-0000-5061-7138}\inst{\ref{aff171}}
\and G.~Leroy\orcid{0009-0004-2523-4425}\inst{\ref{aff23},\ref{aff24}}
\and G.~F.~Lesci\orcid{0000-0002-4607-2830}\inst{\ref{aff14},\ref{aff2}}
\and J.~Lesgourgues\orcid{0000-0001-7627-353X}\inst{\ref{aff67}}
\and T.~I.~Liaudat\orcid{0000-0002-9104-314X}\inst{\ref{aff172}}
\and S.~J.~Liu\orcid{0000-0001-7680-2139}\inst{\ref{aff85}}
\and A.~Loureiro\orcid{0000-0002-4371-0876}\inst{\ref{aff173},\ref{aff174}}
\and J.~Macias-Perez\orcid{0000-0002-5385-2763}\inst{\ref{aff175}}
\and G.~Maggio\orcid{0000-0003-4020-4836}\inst{\ref{aff48}}
\and M.~Magliocchetti\orcid{0000-0001-9158-4838}\inst{\ref{aff85}}
\and F.~Mannucci\orcid{0000-0002-4803-2381}\inst{\ref{aff176}}
\and R.~Maoli\orcid{0000-0002-6065-3025}\inst{\ref{aff177},\ref{aff70}}
\and C.~J.~A.~P.~Martins\orcid{0000-0002-4886-9261}\inst{\ref{aff178},\ref{aff58}}
\and L.~Maurin\orcid{0000-0002-8406-0857}\inst{\ref{aff41}}
\and M.~Migliaccio\inst{\ref{aff179},\ref{aff180}}
\and M.~Miluzio\inst{\ref{aff42},\ref{aff181}}
\and P.~Monaco\orcid{0000-0003-2083-7564}\inst{\ref{aff154},\ref{aff48},\ref{aff49},\ref{aff47}}
\and C.~Moretti\orcid{0000-0003-3314-8936}\inst{\ref{aff50},\ref{aff145},\ref{aff48},\ref{aff47},\ref{aff49}}
\and G.~Morgante\inst{\ref{aff2}}
\and C.~Murray\inst{\ref{aff110}}
\and S.~Nadathur\orcid{0000-0001-9070-3102}\inst{\ref{aff19}}
\and K.~Naidoo\orcid{0000-0002-9182-1802}\inst{\ref{aff19}}
\and A.~Navarro-Alsina\orcid{0000-0002-3173-2592}\inst{\ref{aff107}}
\and S.~Nesseris\orcid{0000-0002-0567-0324}\inst{\ref{aff146}}
\and F.~Passalacqua\orcid{0000-0002-8606-4093}\inst{\ref{aff124},\ref{aff86}}
\and K.~Paterson\orcid{0000-0001-8340-3486}\inst{\ref{aff21}}
\and L.~Patrizii\inst{\ref{aff3}}
\and A.~Pisani\orcid{0000-0002-6146-4437}\inst{\ref{aff87},\ref{aff182}}
\and D.~Potter\orcid{0000-0002-0757-5195}\inst{\ref{aff171}}
\and S.~Quai\orcid{0000-0002-0449-8163}\inst{\ref{aff14},\ref{aff2}}
\and M.~Radovich\orcid{0000-0002-3585-866X}\inst{\ref{aff51}}
\and P.~Reimberg\orcid{0000-0003-3410-0280}\inst{\ref{aff112}}
\and P.-F.~Rocci\inst{\ref{aff41}}
\and G.~Rodighiero\orcid{0000-0002-9415-2296}\inst{\ref{aff124},\ref{aff51}}
\and S.~Sacquegna\orcid{0000-0002-8433-6630}\inst{\ref{aff139},\ref{aff140},\ref{aff141}}
\and M.~Sahl\'en\orcid{0000-0003-0973-4804}\inst{\ref{aff183}}
\and D.~B.~Sanders\orcid{0000-0002-1233-9998}\inst{\ref{aff72}}
\and E.~Sarpa\orcid{0000-0002-1256-655X}\inst{\ref{aff50},\ref{aff145},\ref{aff49}}
\and A.~Schneider\orcid{0000-0001-7055-8104}\inst{\ref{aff171}}
\and M.~Schultheis\inst{\ref{aff109}}
\and D.~Sciotti\orcid{0009-0008-4519-2620}\inst{\ref{aff70},\ref{aff108}}
\and E.~Sellentin\inst{\ref{aff184},\ref{aff64}}
\and F.~Shankar\orcid{0000-0001-8973-5051}\inst{\ref{aff185}}
\and L.~C.~Smith\orcid{0000-0002-3259-2771}\inst{\ref{aff186}}
\and S.~A.~Stanford\orcid{0000-0003-0122-0841}\inst{\ref{aff187}}
\and K.~Tanidis\orcid{0000-0001-9843-5130}\inst{\ref{aff137}}
\and C.~Tao\orcid{0000-0001-7961-8177}\inst{\ref{aff87}}
\and G.~Testera\inst{\ref{aff56}}
\and R.~Teyssier\orcid{0000-0001-7689-0933}\inst{\ref{aff182}}
\and S.~Tosi\orcid{0000-0002-7275-9193}\inst{\ref{aff55},\ref{aff56},\ref{aff45}}
\and A.~Troja\orcid{0000-0003-0239-4595}\inst{\ref{aff124},\ref{aff86}}
\and M.~Tucci\inst{\ref{aff84}}
\and C.~Valieri\inst{\ref{aff3}}
\and A.~Venhola\orcid{0000-0001-6071-4564}\inst{\ref{aff188}}
\and D.~Vergani\orcid{0000-0003-0898-2216}\inst{\ref{aff2}}
\and G.~Verza\orcid{0000-0002-1886-8348}\inst{\ref{aff189}}
\and P.~Vielzeuf\orcid{0000-0003-2035-9339}\inst{\ref{aff87}}
\and N.~A.~Walton\orcid{0000-0003-3983-8778}\inst{\ref{aff186}}
\and E.~Soubrie\orcid{0000-0001-9295-1863}\inst{\ref{aff41}}
\and D.~Scott\orcid{0000-0002-6878-9840}\inst{\ref{aff190}}}
										   
\institute{Dipartimento di Fisica "Aldo Pontremoli", Universit\`a degli Studi di Milano, Via Celoria 16, 20133 Milano, Italy\label{aff1}
\and
INAF-Osservatorio di Astrofisica e Scienza dello Spazio di Bologna, Via Piero Gobetti 93/3, 40129 Bologna, Italy\label{aff2}
\and
INFN-Sezione di Bologna, Viale Berti Pichat 6/2, 40127 Bologna, Italy\label{aff3}
\and
Instituto de F\'isica de Cantabria, Edificio Juan Jord\'a, Avenida de los Castros, 39005 Santander, Spain\label{aff4}
\and
Institute of Physics, Laboratory of Astrophysics, Ecole Polytechnique F\'ed\'erale de Lausanne (EPFL), Observatoire de Sauverny, 1290 Versoix, Switzerland\label{aff5}
\and
SCITAS, Ecole Polytechnique F\'ed\'erale de Lausanne (EPFL), 1015 Lausanne, Switzerland\label{aff6}
\and
INAF-IASF Milano, Via Alfonso Corti 12, 20133 Milano, Italy\label{aff7}
\and
Dipartimento di Fisica e Scienze della Terra, Universit\`a degli Studi di Ferrara, Via Giuseppe Saragat 1, 44122 Ferrara, Italy\label{aff8}
\and
INAF-Osservatorio Astronomico di Capodimonte, Via Moiariello 16, 80131 Napoli, Italy\label{aff9}
\and
Institut de Recherche en Astrophysique et Plan\'etologie (IRAP), Universit\'e de Toulouse, CNRS, UPS, CNES, 14 Av. Edouard Belin, 31400 Toulouse, France\label{aff10}
\and
Minnesota Institute for Astrophysics, University of Minnesota, 116 Church St SE, Minneapolis, MN 55455, USA\label{aff11}
\and
Kapteyn Astronomical Institute, University of Groningen, PO Box 800, 9700 AV Groningen, The Netherlands\label{aff12}
\and
Department of Physics \& Astronomy, University of California Irvine, Irvine CA 92697, USA\label{aff13}
\and
Dipartimento di Fisica e Astronomia "Augusto Righi" - Alma Mater Studiorum Universit\`a di Bologna, via Piero Gobetti 93/2, 40129 Bologna, Italy\label{aff14}
\and
Technical University of Munich, TUM School of Natural Sciences, Physics Department, James-Franck-Str.~1, 85748 Garching, Germany\label{aff15}
\and
Max-Planck-Institut f\"ur Astrophysik, Karl-Schwarzschild-Str.~1, 85748 Garching, Germany\label{aff16}
\and
Aix-Marseille Universit\'e, CNRS, CNES, LAM, Marseille, France\label{aff17}
\and
Institut d'Astrophysique de Paris, UMR 7095, CNRS, and Sorbonne Universit\'e, 98 bis boulevard Arago, 75014 Paris, France\label{aff18}
\and
Institute of Cosmology and Gravitation, University of Portsmouth, Portsmouth PO1 3FX, UK\label{aff19}
\and
Laboratoire univers et particules de Montpellier, Universit\'e de Montpellier, CNRS, 34090 Montpellier, France\label{aff20}
\and
Max-Planck-Institut f\"ur Astronomie, K\"onigstuhl 17, 69117 Heidelberg, Germany\label{aff21}
\and
STAR Institute, University of Li{\`e}ge, Quartier Agora, All\'ee du six Ao\^ut 19c, 4000 Li\`ege, Belgium\label{aff22}
\and
Department of Physics, Centre for Extragalactic Astronomy, Durham University, South Road, Durham, DH1 3LE, UK\label{aff23}
\and
Department of Physics, Institute for Computational Cosmology, Durham University, South Road, Durham, DH1 3LE, UK\label{aff24}
\and
Departamento F\'isica Aplicada, Universidad Polit\'ecnica de Cartagena, Campus Muralla del Mar, 30202 Cartagena, Murcia, Spain\label{aff25}
\and
Center for Frontier Science, Chiba University, 1-33 Yayoi-cho, Inage-ku, Chiba 263-8522, Japan\label{aff26}
\and
Department of Physics, Graduate School of Science, Chiba University, 1-33 Yayoi-Cho, Inage-Ku, Chiba 263-8522, Japan\label{aff27}
\and
Dipartimento di Fisica e Astronomia, Universit\`a di Bologna, Via Gobetti 93/2, 40129 Bologna, Italy\label{aff28}
\and
Centre de Recherche Astrophysique de Lyon, UMR5574, CNRS, Universit\'e Claude Bernard Lyon 1, ENS de Lyon, 69230, Saint-Genis-Laval, France\label{aff29}
\and
University of Applied Sciences and Arts of Northwestern Switzerland, School of Engineering, 5210 Windisch, Switzerland\label{aff30}
\and
Universit\"ats-Sternwarte M\"unchen, Fakult\"at f\"ur Physik, Ludwig-Maximilians-Universit\"at M\"unchen, Scheinerstrasse 1, 81679 M\"unchen, Germany\label{aff31}
\and
Max Planck Institute for Extraterrestrial Physics, Giessenbachstr. 1, 85748 Garching, Germany\label{aff32}
\and
School of Physics and Astronomy, University of Birmingham, Birmingham, B15 2TT, UK\label{aff33}
\and
Department of Astrophysics, University of Vienna, T\"urkenschanzstrasse 17, 1180 Vienna, Austria\label{aff34}
\and
Department of Physics and Astronomy, Lehman College of the CUNY, Bronx, NY 10468, USA\label{aff35}
\and
American Museum of Natural History, Department of Astrophysics, New York, NY 10024, USA\label{aff36}
\and
Caltech/IPAC, 1200 E. California Blvd., Pasadena, CA 91125, USA\label{aff37}
\and
Institut de Ci\`{e}ncies del Cosmos (ICCUB), Universitat de Barcelona (IEEC-UB), Mart\'{i} i Franqu\`{e}s 1, 08028 Barcelona, Spain\label{aff38}
\and
Shanghai Astronomical Observatory (SHAO), Nandan Road 80, Shanghai 200030, China\label{aff39}
\and
Max Planck Institute for Gravitational Physics (Albert Einstein Institute), Am Muhlenberg 1, D-14476 Potsdam-Golm, Germany\label{aff40}
\and
Universit\'e Paris-Saclay, CNRS, Institut d'astrophysique spatiale, 91405, Orsay, France\label{aff41}
\and
ESAC/ESA, Camino Bajo del Castillo, s/n., Urb. Villafranca del Castillo, 28692 Villanueva de la Ca\~nada, Madrid, Spain\label{aff42}
\and
School of Mathematics and Physics, University of Surrey, Guildford, Surrey, GU2 7XH, UK\label{aff43}
\and
Institut f\"ur Theoretische Physik, University of Heidelberg, Philosophenweg 16, 69120 Heidelberg, Germany\label{aff44}
\and
INAF-Osservatorio Astronomico di Brera, Via Brera 28, 20122 Milano, Italy\label{aff45}
\and
Universit\'e Paris-Saclay, Universit\'e Paris Cit\'e, CEA, CNRS, AIM, 91191, Gif-sur-Yvette, France\label{aff46}
\and
IFPU, Institute for Fundamental Physics of the Universe, via Beirut 2, 34151 Trieste, Italy\label{aff47}
\and
INAF-Osservatorio Astronomico di Trieste, Via G. B. Tiepolo 11, 34143 Trieste, Italy\label{aff48}
\and
INFN, Sezione di Trieste, Via Valerio 2, 34127 Trieste TS, Italy\label{aff49}
\and
SISSA, International School for Advanced Studies, Via Bonomea 265, 34136 Trieste TS, Italy\label{aff50}
\and
INAF-Osservatorio Astronomico di Padova, Via dell'Osservatorio 5, 35122 Padova, Italy\label{aff51}
\and
Centre National d'Etudes Spatiales -- Centre spatial de Toulouse, 18 avenue Edouard Belin, 31401 Toulouse Cedex 9, France\label{aff52}
\and
Space Science Data Center, Italian Space Agency, via del Politecnico snc, 00133 Roma, Italy\label{aff53}
\and
INAF-Osservatorio Astrofisico di Torino, Via Osservatorio 20, 10025 Pino Torinese (TO), Italy\label{aff54}
\and
Dipartimento di Fisica, Universit\`a di Genova, Via Dodecaneso 33, 16146, Genova, Italy\label{aff55}
\and
INFN-Sezione di Genova, Via Dodecaneso 33, 16146, Genova, Italy\label{aff56}
\and
Department of Physics "E. Pancini", University Federico II, Via Cinthia 6, 80126, Napoli, Italy\label{aff57}
\and
Instituto de Astrof\'isica e Ci\^encias do Espa\c{c}o, Universidade do Porto, CAUP, Rua das Estrelas, PT4150-762 Porto, Portugal\label{aff58}
\and
Faculdade de Ci\^encias da Universidade do Porto, Rua do Campo de Alegre, 4150-007 Porto, Portugal\label{aff59}
\and
Dipartimento di Fisica, Universit\`a degli Studi di Torino, Via P. Giuria 1, 10125 Torino, Italy\label{aff60}
\and
INFN-Sezione di Torino, Via P. Giuria 1, 10125 Torino, Italy\label{aff61}
\and
European Space Agency/ESTEC, Keplerlaan 1, 2201 AZ Noordwijk, The Netherlands\label{aff62}
\and
Institute Lorentz, Leiden University, Niels Bohrweg 2, 2333 CA Leiden, The Netherlands\label{aff63}
\and
Leiden Observatory, Leiden University, Einsteinweg 55, 2333 CC Leiden, The Netherlands\label{aff64}
\and
Centro de Investigaciones Energ\'eticas, Medioambientales y Tecnol\'ogicas (CIEMAT), Avenida Complutense 40, 28040 Madrid, Spain\label{aff65}
\and
Port d'Informaci\'{o} Cient\'{i}fica, Campus UAB, C. Albareda s/n, 08193 Bellaterra (Barcelona), Spain\label{aff66}
\and
Institute for Theoretical Particle Physics and Cosmology (TTK), RWTH Aachen University, 52056 Aachen, Germany\label{aff67}
\and
Institute of Space Sciences (ICE, CSIC), Campus UAB, Carrer de Can Magrans, s/n, 08193 Barcelona, Spain\label{aff68}
\and
Institut d'Estudis Espacials de Catalunya (IEEC),  Edifici RDIT, Campus UPC, 08860 Castelldefels, Barcelona, Spain\label{aff69}
\and
INAF-Osservatorio Astronomico di Roma, Via Frascati 33, 00078 Monteporzio Catone, Italy\label{aff70}
\and
INFN section of Naples, Via Cinthia 6, 80126, Napoli, Italy\label{aff71}
\and
Institute for Astronomy, University of Hawaii, 2680 Woodlawn Drive, Honolulu, HI 96822, USA\label{aff72}
\and
Dipartimento di Fisica e Astronomia "Augusto Righi" - Alma Mater Studiorum Universit\`a di Bologna, Viale Berti Pichat 6/2, 40127 Bologna, Italy\label{aff73}
\and
Instituto de Astrof\'{\i}sica de Canarias, V\'{\i}a L\'actea, 38205 La Laguna, Tenerife, Spain\label{aff74}
\and
Institute for Astronomy, University of Edinburgh, Royal Observatory, Blackford Hill, Edinburgh EH9 3HJ, UK\label{aff75}
\and
Jodrell Bank Centre for Astrophysics, Department of Physics and Astronomy, University of Manchester, Oxford Road, Manchester M13 9PL, UK\label{aff76}
\and
European Space Agency/ESRIN, Largo Galileo Galilei 1, 00044 Frascati, Roma, Italy\label{aff77}
\and
Universit\'e Claude Bernard Lyon 1, CNRS/IN2P3, IP2I Lyon, UMR 5822, Villeurbanne, F-69100, France\label{aff78}
\and
Instituci\'o Catalana de Recerca i Estudis Avan\c{c}ats (ICREA), Passeig de Llu\'{\i}s Companys 23, 08010 Barcelona, Spain\label{aff79}
\and
UCB Lyon 1, CNRS/IN2P3, IUF, IP2I Lyon, 4 rue Enrico Fermi, 69622 Villeurbanne, France\label{aff80}
\and
Mullard Space Science Laboratory, University College London, Holmbury St Mary, Dorking, Surrey RH5 6NT, UK\label{aff81}
\and
Departamento de F\'isica, Faculdade de Ci\^encias, Universidade de Lisboa, Edif\'icio C8, Campo Grande, PT1749-016 Lisboa, Portugal\label{aff82}
\and
Instituto de Astrof\'isica e Ci\^encias do Espa\c{c}o, Faculdade de Ci\^encias, Universidade de Lisboa, Campo Grande, 1749-016 Lisboa, Portugal\label{aff83}
\and
Department of Astronomy, University of Geneva, ch. d'Ecogia 16, 1290 Versoix, Switzerland\label{aff84}
\and
INAF-Istituto di Astrofisica e Planetologia Spaziali, via del Fosso del Cavaliere, 100, 00100 Roma, Italy\label{aff85}
\and
INFN-Padova, Via Marzolo 8, 35131 Padova, Italy\label{aff86}
\and
Aix-Marseille Universit\'e, CNRS/IN2P3, CPPM, Marseille, France\label{aff87}
\and
INFN-Bologna, Via Irnerio 46, 40126 Bologna, Italy\label{aff88}
\and
School of Physics, HH Wills Physics Laboratory, University of Bristol, Tyndall Avenue, Bristol, BS8 1TL, UK\label{aff89}
\and
FRACTAL S.L.N.E., calle Tulip\'an 2, Portal 13 1A, 28231, Las Rozas de Madrid, Spain\label{aff90}
\and
INFN-Sezione di Milano, Via Celoria 16, 20133 Milano, Italy\label{aff91}
\and
Institute of Theoretical Astrophysics, University of Oslo, P.O. Box 1029 Blindern, 0315 Oslo, Norway\label{aff92}
\and
Jet Propulsion Laboratory, California Institute of Technology, 4800 Oak Grove Drive, Pasadena, CA, 91109, USA\label{aff93}
\and
Felix Hormuth Engineering, Goethestr. 17, 69181 Leimen, Germany\label{aff94}
\and
Technical University of Denmark, Elektrovej 327, 2800 Kgs. Lyngby, Denmark\label{aff95}
\and
Cosmic Dawn Center (DAWN), Denmark\label{aff96}
\and
NASA Goddard Space Flight Center, Greenbelt, MD 20771, USA\label{aff97}
\and
Department of Physics and Astronomy, University College London, Gower Street, London WC1E 6BT, UK\label{aff98}
\and
Department of Physics and Helsinki Institute of Physics, Gustaf H\"allstr\"omin katu 2, 00014 University of Helsinki, Finland\label{aff99}
\and
Universit\'e de Gen\`eve, D\'epartement de Physique Th\'eorique and Centre for Astroparticle Physics, 24 quai Ernest-Ansermet, CH-1211 Gen\`eve 4, Switzerland\label{aff100}
\and
Department of Physics, P.O. Box 64, 00014 University of Helsinki, Finland\label{aff101}
\and
Helsinki Institute of Physics, Gustaf H{\"a}llstr{\"o}min katu 2, University of Helsinki, Helsinki, Finland\label{aff102}
\and
Centre de Calcul de l'IN2P3/CNRS, 21 avenue Pierre de Coubertin 69627 Villeurbanne Cedex, France\label{aff103}
\and
Laboratoire d'etude de l'Univers et des phenomenes eXtremes, Observatoire de Paris, Universit\'e PSL, Sorbonne Universit\'e, CNRS, 92190 Meudon, France\label{aff104}
\and
SKA Observatory, Jodrell Bank, Lower Withington, Macclesfield, Cheshire SK11 9FT, UK\label{aff105}
\and
University of Applied Sciences and Arts of Northwestern Switzerland, School of Computer Science, 5210 Windisch, Switzerland\label{aff106}
\and
Universit\"at Bonn, Argelander-Institut f\"ur Astronomie, Auf dem H\"ugel 71, 53121 Bonn, Germany\label{aff107}
\and
INFN-Sezione di Roma, Piazzale Aldo Moro, 2 - c/o Dipartimento di Fisica, Edificio G. Marconi, 00185 Roma, Italy\label{aff108}
\and
Universit\'e C\^{o}te d'Azur, Observatoire de la C\^{o}te d'Azur, CNRS, Laboratoire Lagrange, Bd de l'Observatoire, CS 34229, 06304 Nice cedex 4, France\label{aff109}
\and
Universit\'e Paris Cit\'e, CNRS, Astroparticule et Cosmologie, 75013 Paris, France\label{aff110}
\and
CNRS-UCB International Research Laboratory, Centre Pierre Binetruy, IRL2007, CPB-IN2P3, Berkeley, USA\label{aff111}
\and
Institut d'Astrophysique de Paris, 98bis Boulevard Arago, 75014, Paris, France\label{aff112}
\and
Aurora Technology for European Space Agency (ESA), Camino bajo del Castillo, s/n, Urbanizacion Villafranca del Castillo, Villanueva de la Ca\~nada, 28692 Madrid, Spain\label{aff113}
\and
OCA, P.H.C Boulevard de l'Observatoire CS 34229, 06304 Nice Cedex 4, France\label{aff114}
\and
Institut de F\'{i}sica d'Altes Energies (IFAE), The Barcelona Institute of Science and Technology, Campus UAB, 08193 Bellaterra (Barcelona), Spain\label{aff115}
\and
School of Mathematics, Statistics and Physics, Newcastle University, Herschel Building, Newcastle-upon-Tyne, NE1 7RU, UK\label{aff116}
\and
DARK, Niels Bohr Institute, University of Copenhagen, Jagtvej 155, 2200 Copenhagen, Denmark\label{aff117}
\and
Waterloo Centre for Astrophysics, University of Waterloo, Waterloo, Ontario N2L 3G1, Canada\label{aff118}
\and
Department of Physics and Astronomy, University of Waterloo, Waterloo, Ontario N2L 3G1, Canada\label{aff119}
\and
Perimeter Institute for Theoretical Physics, Waterloo, Ontario N2L 2Y5, Canada\label{aff120}
\and
Institute of Space Science, Str. Atomistilor, nr. 409 M\u{a}gurele, Ilfov, 077125, Romania\label{aff121}
\and
Consejo Superior de Investigaciones Cientificas, Calle Serrano 117, 28006 Madrid, Spain\label{aff122}
\and
Universidad de La Laguna, Departamento de Astrof\'{\i}sica, 38206 La Laguna, Tenerife, Spain\label{aff123}
\and
Dipartimento di Fisica e Astronomia "G. Galilei", Universit\`a di Padova, Via Marzolo 8, 35131 Padova, Italy\label{aff124}
\and
Universit\'e St Joseph; Faculty of Sciences, Beirut, Lebanon\label{aff125}
\and
Departamento de F\'isica, FCFM, Universidad de Chile, Blanco Encalada 2008, Santiago, Chile\label{aff126}
\and
Satlantis, University Science Park, Sede Bld 48940, Leioa-Bilbao, Spain\label{aff127}
\and
Department of Physics, Royal Holloway, University of London, TW20 0EX, UK\label{aff128}
\and
Infrared Processing and Analysis Center, California Institute of Technology, Pasadena, CA 91125, USA\label{aff129}
\and
Instituto de Astrof\'isica e Ci\^encias do Espa\c{c}o, Faculdade de Ci\^encias, Universidade de Lisboa, Tapada da Ajuda, 1349-018 Lisboa, Portugal\label{aff130}
\and
Cosmic Dawn Center (DAWN)\label{aff131}
\and
Niels Bohr Institute, University of Copenhagen, Jagtvej 128, 2200 Copenhagen, Denmark\label{aff132}
\and
Universidad Polit\'ecnica de Cartagena, Departamento de Electr\'onica y Tecnolog\'ia de Computadoras,  Plaza del Hospital 1, 30202 Cartagena, Spain\label{aff133}
\and
Centre for Information Technology, University of Groningen, P.O. Box 11044, 9700 CA Groningen, The Netherlands\label{aff134}
\and
Istituto Nazionale di Fisica Nucleare, Sezione di Ferrara, Via Giuseppe Saragat 1, 44122 Ferrara, Italy\label{aff135}
\and
INAF, Istituto di Radioastronomia, Via Piero Gobetti 101, 40129 Bologna, Italy\label{aff136}
\and
Department of Physics, Oxford University, Keble Road, Oxford OX1 3RH, UK\label{aff137}
\and
Zentrum f\"ur Astronomie, Universit\"at Heidelberg, Philosophenweg 12, 69120 Heidelberg, Germany\label{aff138}
\and
Department of Mathematics and Physics E. De Giorgi, University of Salento, Via per Arnesano, CP-I93, 73100, Lecce, Italy\label{aff139}
\and
INFN, Sezione di Lecce, Via per Arnesano, CP-193, 73100, Lecce, Italy\label{aff140}
\and
INAF-Sezione di Lecce, c/o Dipartimento Matematica e Fisica, Via per Arnesano, 73100, Lecce, Italy\label{aff141}
\and
INAF - Osservatorio Astronomico di Brera, via Emilio Bianchi 46, 23807 Merate, Italy\label{aff142}
\and
INAF-Osservatorio Astronomico di Brera, Via Brera 28, 20122 Milano, Italy, and INFN-Sezione di Genova, Via Dodecaneso 33, 16146, Genova, Italy\label{aff143}
\and
ICL, Junia, Universit\'e Catholique de Lille, LITL, 59000 Lille, France\label{aff144}
\and
ICSC - Centro Nazionale di Ricerca in High Performance Computing, Big Data e Quantum Computing, Via Magnanelli 2, Bologna, Italy\label{aff145}
\and
Instituto de F\'isica Te\'orica UAM-CSIC, Campus de Cantoblanco, 28049 Madrid, Spain\label{aff146}
\and
CERCA/ISO, Department of Physics, Case Western Reserve University, 10900 Euclid Avenue, Cleveland, OH 44106, USA\label{aff147}
\and
Laboratoire Univers et Th\'eorie, Observatoire de Paris, Universit\'e PSL, Universit\'e Paris Cit\'e, CNRS, 92190 Meudon, France\label{aff148}
\and
Departamento de F{\'\i}sica Fundamental. Universidad de Salamanca. Plaza de la Merced s/n. 37008 Salamanca, Spain\label{aff149}
\and
Universit\'e de Strasbourg, CNRS, Observatoire astronomique de Strasbourg, UMR 7550, 67000 Strasbourg, France\label{aff150}
\and
Center for Data-Driven Discovery, Kavli IPMU (WPI), UTIAS, The University of Tokyo, Kashiwa, Chiba 277-8583, Japan\label{aff151}
\and
Ludwig-Maximilians-University, Schellingstrasse 4, 80799 Munich, Germany\label{aff152}
\and
Max-Planck-Institut f\"ur Physik, Boltzmannstr. 8, 85748 Garching, Germany\label{aff153}
\and
Dipartimento di Fisica - Sezione di Astronomia, Universit\`a di Trieste, Via Tiepolo 11, 34131 Trieste, Italy\label{aff154}
\and
California Institute of Technology, 1200 E California Blvd, Pasadena, CA 91125, USA\label{aff155}
\and
Instituto de Astrof\'isica de Canarias (IAC); Departamento de Astrof\'isica, Universidad de La Laguna (ULL), 38200, La Laguna, Tenerife, Spain\label{aff156}
\and
CEA Saclay, DFR/IRFU, Service d'Astrophysique, Bat. 709, 91191 Gif-sur-Yvette, France\label{aff157}
\and
Department of Astronomy, University of Florida, Bryant Space Science Center, Gainesville, FL 32611, USA\label{aff158}
\and
Department of Computer Science, Aalto University, PO Box 15400, Espoo, FI-00 076, Finland\label{aff159}
\and
Instituto de Astrof\'\i sica de Canarias, c/ Via Lactea s/n, La Laguna 38200, Spain. Departamento de Astrof\'\i sica de la Universidad de La Laguna, Avda. Francisco Sanchez, La Laguna, 38200, Spain\label{aff160}
\and
Ruhr University Bochum, Faculty of Physics and Astronomy, Astronomical Institute (AIRUB), German Centre for Cosmological Lensing (GCCL), 44780 Bochum, Germany\label{aff161}
\and
Astrophysics Research Centre, University of KwaZulu-Natal, Westville Campus, Durban 4041, South Africa\label{aff162}
\and
School of Mathematics, Statistics \& Computer Science, University of KwaZulu-Natal, Westville Campus, Durban 4041, South Africa\label{aff163}
\and
Department of Physics and Astronomy, Vesilinnantie 5, 20014 University of Turku, Finland\label{aff164}
\and
Serco for European Space Agency (ESA), Camino bajo del Castillo, s/n, Urbanizacion Villafranca del Castillo, Villanueva de la Ca\~nada, 28692 Madrid, Spain\label{aff165}
\and
ARC Centre of Excellence for Dark Matter Particle Physics, Melbourne, Australia\label{aff166}
\and
Centre for Astrophysics \& Supercomputing, Swinburne University of Technology,  Hawthorn, Victoria 3122, Australia\label{aff167}
\and
Department of Physics and Astronomy, University of the Western Cape, Bellville, Cape Town, 7535, South Africa\label{aff168}
\and
DAMTP, Centre for Mathematical Sciences, Wilberforce Road, Cambridge CB3 0WA, UK\label{aff169}
\and
Kavli Institute for Cosmology Cambridge, Madingley Road, Cambridge, CB3 0HA, UK\label{aff170}
\and
Department of Astrophysics, University of Zurich, Winterthurerstrasse 190, 8057 Zurich, Switzerland\label{aff171}
\and
IRFU, CEA, Universit\'e Paris-Saclay 91191 Gif-sur-Yvette Cedex, France\label{aff172}
\and
Oskar Klein Centre for Cosmoparticle Physics, Department of Physics, Stockholm University, Stockholm, SE-106 91, Sweden\label{aff173}
\and
Astrophysics Group, Blackett Laboratory, Imperial College London, London SW7 2AZ, UK\label{aff174}
\and
Univ. Grenoble Alpes, CNRS, Grenoble INP, LPSC-IN2P3, 53, Avenue des Martyrs, 38000, Grenoble, France\label{aff175}
\and
INAF-Osservatorio Astrofisico di Arcetri, Largo E. Fermi 5, 50125, Firenze, Italy\label{aff176}
\and
Dipartimento di Fisica, Sapienza Universit\`a di Roma, Piazzale Aldo Moro 2, 00185 Roma, Italy\label{aff177}
\and
Centro de Astrof\'{\i}sica da Universidade do Porto, Rua das Estrelas, 4150-762 Porto, Portugal\label{aff178}
\and
Dipartimento di Fisica, Universit\`a di Roma Tor Vergata, Via della Ricerca Scientifica 1, Roma, Italy\label{aff179}
\and
INFN, Sezione di Roma 2, Via della Ricerca Scientifica 1, Roma, Italy\label{aff180}
\and
HE Space for European Space Agency (ESA), Camino bajo del Castillo, s/n, Urbanizacion Villafranca del Castillo, Villanueva de la Ca\~nada, 28692 Madrid, Spain\label{aff181}
\and
Department of Astrophysical Sciences, Peyton Hall, Princeton University, Princeton, NJ 08544, USA\label{aff182}
\and
Theoretical astrophysics, Department of Physics and Astronomy, Uppsala University, Box 515, 751 20 Uppsala, Sweden\label{aff183}
\and
Mathematical Institute, University of Leiden, Einsteinweg 55, 2333 CA Leiden, The Netherlands\label{aff184}
\and
School of Physics \& Astronomy, University of Southampton, Highfield Campus, Southampton SO17 1BJ, UK\label{aff185}
\and
Institute of Astronomy, University of Cambridge, Madingley Road, Cambridge CB3 0HA, UK\label{aff186}
\and
Department of Physics and Astronomy, University of California, Davis, CA 95616, USA\label{aff187}
\and
Space physics and astronomy research unit, University of Oulu, Pentti Kaiteran katu 1, FI-90014 Oulu, Finland\label{aff188}
\and
Center for Computational Astrophysics, Flatiron Institute, 162 5th Avenue, 10010, New York, NY, USA\label{aff189}
\and
Department of Physics and Astronomy, University of British Columbia, Vancouver, BC V6T 1Z1, Canada\label{aff190}}    